\documentclass[]{jfm}

\usepackage{graphicx}
\usepackage{newtxtext}
\usepackage{newtxmath}
\usepackage{natbib}
\usepackage{hyperref}
\hypersetup{
    colorlinks = true,
    urlcolor   = blue,
    citecolor  = black,
}

\newcommand{\RomanNumeralCaps}[1]
\linenumbers
\usepackage[ruled,boxed]{algorithm2e}
\usepackage{makecell}
\usepackage{graphicx}
\usepackage{bm,booktabs,bigstrut}
\usepackage{overpic}
\usepackage{multirow} % Required for multirows
\usepackage{rotating}

\usepackage{tabularx}
\newcolumntype{P}[1]{>{\centering \arraybackslash}p{#1}}
\newcolumntype{L}{X}
\newcolumntype{C}{>{\centering \arraybackslash}X}
\newcolumntype{R}{>{\raggedright \arraybackslash}X}

\newcommand{\layout}[2]{$\mathcal{L}_{\uppercase\expandafter{\romannumeral#1}}^{#2}$}
\newcommand{\fig}[1]{(\textit{#1})}

\usepackage[x11names]{xcolor}

\title{Deep reinforcement transfer learning for active flow control of a 3D square cylinder under state dimension mismatch}

\author{
    Lei~Yan\aff{1} ,
    Gang~Hu\aff{1,2} \corresp{\email{hugang@hit.edu.cn}},
    Wenli~Chen\aff{3},
    \and Bernd~R.~Noack\aff{4}
}

\affiliation{
    \aff{1}Artificial Intelligence for Wind Engineering (AIWE) Lab, School of Civil and Environmental Engineering, Harbin Institute of Technology, Shenzhen 518055, China
    \aff{2}Guangdong Provincial Key Laboratory of Intelligent and Resilient Structures for Civil Engineering, Harbin Institute of Technology, Shenzhen, 518055, China
    \aff{3}Key Laboratory of Smart Prevention and Mitigation of Civil Engineering Disasters, the Ministry of Industry and Information Technology, Harbin Institute of Technology, Harbin, 150090, China
    \aff{4}School of Mechanical Engineering and Automation, Harbin Institute of Technology, Shenzhen 518055, China
}

\begin{document}
\maketitle

\begin{abstract}
% 本文的重点是找到一种闭环策略，以减少高雷诺数条件下方柱的气动力。利用位于方柱四角的四个射流作为执行器，方柱表面压力传感器作为反馈观测器，实施了深度强化学习算法，以发现有效的控制方案。
% 另外，使用跨域迁移学习解决状态空间不匹配的问题。为了处理这种不匹配，我们制作了嵌入式编码，可以系统地从source网络策略和价值网络工程中提取知识边缘，并将其融合到current的网络中。并与现有SAC算法相融合，命名为state dimensional dismatch(SDTL-SAC)方法。我们首先使用低雷诺数下的2D方柱流场对 DRL 代理进行预训练，再迁移到3D的高雷诺数方柱流场中进行训练。
% 结果表明，我们利用在2D到3D（雷诺数为22000）方柱的迁移任务，展示了不同状态空间的情况下成功的迁移学习。并提升了学习速度。并且DRL控制策略可将阻力降低52.3\%，升力波动被大幅度抑制. 这些结果进一步突出了强化学习在主动流量控制方面优化的可能性，并为在实验或工业系统中高效、稳健和实用地实施这些控制技术铺平了道路。
This paper focuses on developing a deep reinforcement learning (DRL) control strategy to mitigate aerodynamic forces acting on a three dimensional (3D) square cylinder under high Reynolds number flow conditions. Four jets situated at the corners of the square cylinder are used as actuators and pressure probes on the cylinder surface are employed as feedback observers. The Soft Actor-Critic (SAC) algorithm is deployed to identify an effective control scheme. Additionally, we pre-train the DRL agent using a two dimensional (2D) square cylinder flow field at a low Reynolds number ($Re =1000$), followed by transferring it to the 3D square cylinder at $Re =22000$. To address the issue of state dimension mismatch in transfer learning from 2D to 3D case, a state dimension mismatch transfer learning method is developed to enhance the SAC algorithm, named SDTL-SAC. The results demonstrate transfer learning across different state spaces achieves the same control policy as the SAC algorithm, resulting in a significant improvement in training speed with a training cost reduction of 51.1\%. Furthermore, the SAC control strategy leads to a notable 52.3\% reduction in drag coefficient, accompanied by substantial suppression of lift fluctuations. These outcomes underscore the potential of DRL in active flow control, laying the groundwork for efficient, robust, and practical implementation of this control technique in practical engineering.

\end{abstract}

\begin{keywords}
Authors should not enter keywords on the manuscript, as these must be chosen by the author during the online submission process and will then be added during the typesetting process (see \href{https://www.cambridge.org/core/journals/journal-of-fluid-mechanics/information/list-of-keywords}{Keyword PDF} for the full list).  Other classifications will be added at the same time.
\end{keywords}

% {\bf MSC Codes }  {\it(Optional)} Please enter your MSC Codes here

\section{Introduction}\label{sec:intro}
%写方柱流动控制领域               重新写下
% 在流体动力学和风能工程领域，减轻方形圆柱体在高湍流环境下的空气动力是一个关键的研究课题。在高雷诺数条件下，方形圆柱体的尾流经常表现出不稳定性，导致涡流脱落增强，进而加剧阻力，同时引起升力波动。\citet{hu2022attenuation}用数值方法探讨了不稳定风载荷对高层建筑的影响，研究了涡流脱落模式和大气边界层的影响。设计了两种反馈控制策略，利用单个建筑墙体上的压力传感来减缓空气动力侧向力波动，其中线性控制器在模拟中表现出更优越的性能。因此，在高湍流环境中优化方形圆柱体的空气动力特性往往需要使用传感器和反馈控制。这包括实时调整主动流量控制（AFC）策略，以提高方形气缸的气动性能（citep{gao2021active,zhou2005suppression,hu2023attenuation}）。
In the fields of fluid dynamics and wind engineering, mitigating the aerodynamic forces acting on a square cylinder under high turbulence flow environment is a pivotal research topic. Under high Reynolds number flow conditions, the wake of the square cylinder exhibits instability, leading to complex vortex shedding. Flow control to mitigate aerodynamic forces of square cylinders is crucial to ensure their aerodynamic stability. In the past, a number of passive and active flow control (AFC) strategies have been developed. Compared to the passive flow control, the AFC strategies are normally more efficient. 

%写方柱在2D DRL流动控制领域的研究     这里再加点内容
% 流量控制已开始不断融合先进的机器学习方法，其中一个显著的重点是利用植根于试错概念的 DRL 范式。以往的研究已经令人信服地证明了 DRL 在获取复杂环境控制策略方面的功效，尤其是那些以高维和非线性为特征的环境。DRL 范式利用评估数据来调整控制法则，旨在最大限度地提高指定的性能指标。其基本概念是促进有利于目标指标的行动或政策。在 DRL 算法中，网络结构的任务是制定策略，其作用是根据观察到的环境数据生成控制行动。因此，探索新的控制策略对于在不同环境中实现最优控制性能至关重要。不断扩展的文献
Active flow control (AFC) has embraced the continuous integration of advanced machine learning methods, with a notable emphasis on leveraging the deep reinforcement learning (DRL) paradigm rooted in trial-and-error concepts. The DRL paradigm utilizes evaluation data to adapt the control law, aiming to maximize a specified performance metric. The fundamental concept involves promoting actions or policies conducive to the targeted metric. Currently, there is substantial relevant research employing DRL to address issues related to AFC including vortex-induced vibration \citep{zheng2021active, ren2021bluff, chen2023deep}, shape optimization \citep{liu2022prediction, viquerat2021direct, li2021knowledge}, optimization of the movement collective fish \citep{gazzola2016learning}, and turbulent flow control \citep{guastoni2023deep, ren2019active,yan2023stabilizing}. The application of DRL for reducing the drag of the cylinder was first introduced by \citet{Rabault2019Artificial}, resulting in a notable stable drag reduction of approximately 8\% at $Re$ = 100. Meanwhile, \citet{tang2020robust} implemented four synthetic jets symmetrically positioned on the lower and upper sides of a cylinder to control the flow, as well as demonstrated that DRL can effectively discern various control strategies across different $Re$. \citet{han2022deep} minimized the mean drag coefficient of a cylinder by optimizing the angular velocity of a rotating circle. Previous studies have convincingly demonstrated the efficacy of DRL in acquiring control policies for complex environments, particularly those characterized by high dimensionality and nonlinearity \citep{vignon2023recent}.

%找五六篇文献  %真实
% 2D到3D   现有迁移学习   引到这篇文章
% 然而，尽管DRL算法在二维流动控制在理解和改善流场行为方面发挥了关键作用\citep{paris2021robust,hamalainen2020ppo,paris2023reinforcement,tang2020robust}，二维流动控制在模拟真实流场中的复杂三维特性上仍然存在一定的局限性。实际上，流场控制通常涉及多个维度的相互作用，而简单的二维模型往往无法准确捕捉到这些相互作用的本质。因此，为了更全面地理解和掌握流动行为，研究人员逐渐将焦点从二维流动控制扩展到更为复杂的三维流动控制\citep{paris2021robust}。即便是在三维流动控制的框架下，由于流场的动态复杂性和非线性特性，仍然面临训练控制策略费时等困难。例如，在 \citet{fan2020reinforcement}的研究中，在 $Re = 10^4$ 的情况下，需要一个月的时间来训练 DRL 控制策略。当 $Re$ 上升到更高水平时，与直接 DRL 训练相关的时间成本就变得不可接受了。
% 为了克服这些挑战，引入迁移学习成为一种可行的方法。迁移学习通过在不同的任务或领域中共享知识，从而加速模型在目标任务上的学习过程，提高了对复杂流场的建模和控制的准确性。% \citet{he2023policy}介绍了一种利用DRL的策略转移战略，展示了在二维和三维流动环境中减少阻力的效果，在三维环境中，当Re$越高，效果越好。虽然可以从模拟数据中学习，但模拟速度会带来限制。然而，citet{wang2023deep}的后续工作展示了迁移学习在高Re$条件下加速DRL控制气缸流的应用，实现了稳定性的增强和训练次数的减少，并特别关注了Re = 1.4 \times 10^5$条件下的唤醒流。这意味着从低Re$到高Re$的知识迁移是可行的。
% 上述许多研究主要依靠直接迁移学习，利用状态空间一致性进行加速。然而，适用于三维流场控制的跨域迁移学习算法却明显缺乏。具体来说，由于状态空间的变化，将控制策略从二维流场迁移到三维流场环境是一个巨大的挑战。传统的迁移学习方法难以解决这种状态空间变化引起的不一致性，阻碍了 DRL 代理从不同的状态空间流场策略中有效地学习知识。认识到这一挑战后，我们在 SAC 算法框架内引入了状态空间不匹配的考虑因素，并提出了 SDTL-SAC 方法。在圆柱体周围流动的背景下，该方法通过编码器促进跨域转移学习，并从二维流场中提取控制策略，从而大大提高了 DRL 算法的收敛性能。
Despite the crucial role played by DRL algorithms in understanding and improving the behavior of 2D flow control \citep{paris2021robust, paris2023reinforcement}, there are inherent limitations in 2D flow control when simulating the intricate 3D characteristics of real flow fields. In practice, flow field control often involves interactions across multiple dimensions, and 2D models frequently fail to accurately capture the essence of these interactions. Consequently, to achieve a more comprehensive understanding and mastery of flow behavior, researchers have progressively shifted their focus from 2D flow control to the more intricate realm of 3D flow control \citep{suarez2023active}. However, within the framework of 3D flow control, challenges persist due to the dynamic complexity and non-linear nature of flow fields, leading to difficulties such as time consuming training of control policy. For instance, in the investigation by \citet{fan2020reinforcement}, a month is required to train the DRL control strategy with $Re = 10^4$. As $Re$ escalates to higher levels, the time cost associated with direct DRL training becomes unacceptable. 

To address these challenges, the introduction of transfer learning has emerged as a viable approach. transfer learning facilitates the acceleration of the learning process for models in target tasks by sharing knowledge across various tasks or domains. This practice enhances the accuracy of modeling and control for complex flow fields. \citet{he2023policy} introduces a policy transfer strategy utilizing DRL, demonstrating drag reduction in both 2D and 3D flow environments, with increased effectiveness at higher $Re$ in the 3D environment. While learning from simulated data is uninhibited, the speed of simulation imposes limitations. Moreover, subsequent work by \citet{wang2023deep} showcases the application of transfer learning to expedite DRL in controlling cylinder flows at high $Re$, achieving enhanced stability and reduced training episodes, with a specific focus on wake flow at $Re = 1.4 \times 10^5$. This implies the feasibility of transferring knowledge from low $Re$ to high $Re$.

%补充连接性
The majority of the aforementioned studies primarily rely on direct transfer learning under state dimension consistency for acceleration. However, there is a notable absence of a cross-domain transfer learning algorithm applicable to 3D flow control. One possible reason is that transferring a control strategy from a 2D flow field to a 3D flow field environment presents a significant challenge due to changes in the state space. Conventional transfer learning methods struggle to address the inconsistency induced by this state-space shift, impeding DRL agents from effectively learning knowledge from diverse state-space flow field policy. Acknowledging this challenge, the state dimension mismatch transfer learning coupled with the SAC algorithm (SDTL-SAC) is proposed in this study. In the 3D flow around a square cylinder, this method substantially enhances the convergence performance of the DRL algorithm by facilitating cross-domain transfer learning through an encoder and extracting control policy from 2D flow fields.

%最后总结
% 之前迁移学习研究大都依赖源策略与当前策略状态空间对齐。在当前的研究中，我们旨在引入一种在状态空间不匹配时进行跨域迁移学习的方法。据我们所知，在利用DRL进行流量控制的背景下，这个问题从未被解决过。通过选择2D流场迁移到3D方柱流场的测试案例，本文对该算法进行了批判性分析。因此，本文的目标是双重的：介绍第一个解决流动控制中跨域迁移学习这一基本问题的RL算法，并确定未来的挑战和研究方向，为即将开展的相关工作铺平道路。在介绍所提出的基于state dimension mismatch的迁移算法之前，第 2 节介绍了作为算法基础的RL方法。第3节给出了仿真模型、数值方法的细节，第4节将介绍Re=22000下方柱的控制效果，并讨论该方法在测试案例的结果。最后，给出了本文的结论，并讨论了所提出的迁移算法的局限性。 
This study initiates by presenting the control performance within the 3D flow field of a square cylinder employing a multiple jet configuration scheme. A cross-domain transfer learning method is used to transfer the domain knowledge from the 2D square cylinder flow field to the 3D square cylinder flow field under the mismatch in state dimensions. The rest of the paper is organized as follows: Section \ref{sec:numerical} provides details of the simulation model and the numerical method of the 2D and 3D square cylinder. Section \ref{sec:DRL} outlines the DRL methodology underlying the algorithm and introduces the SDTL-SAC. The performance of the DRL control on a square cylinder under $\mathrm{Re} = 22000$ and the results of the transfer learning method are discussed in Section \ref{sec:Results}. Finally, Section \ref{sec:Conclusions} gives the conclusion of the paper and the limitations of the proposed transfer algorithm are discussed.

% Methodology
\section{Description of the flow configuration and numerical methods} \label{sec:numerical}
This section begins with a numerical simulation configuration of a 2D square cylinder in Section \ref{sec:cfd1}. Section \ref{sec:cfd2} provides a comprehensive introduction to the numerical simulation configuration of a 3D square cylinder. The accuracy of these numerical algorithms is then scrutinized for validation purposes. Finally, the flow control configuration scheme is introduced in Section \ref{sec:cfd3}, incorporating the placement of four jets at the corners of the square cylinder as actuators for AFC. The goal is to mitigate the drag force and lift fluctuation of the square cylinder.

\subsection{Numerical simulation configuration of 2D square cylinder}  \label{sec:cfd1}

%加2D计算域，讲两三段
\begin{figure}
\centering
\begin{overpic}[width=0.6\textwidth]{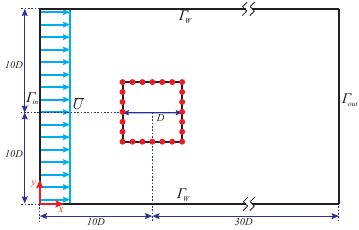}
\end{overpic}
\caption{Description of 2D square cylinder numerical setup: the origin of the coordinates is located at the center of the square cylinder. $\Gamma_{in}$ stands the inflow velocity with a uniform flow profile, while $\Gamma_{out}$ is put for the outflow. Non-slip wall boundary constraint $\Gamma_W$ is applied on the bottom and top of the channel, as well as on the surface of the square cylinder. The red dots represent the positions of the 24 pressure probes.} 
\label{fig:2d_flow_over_square}
\end{figure}

The open-source computational fluid dynamics package OpenFOAM \citep{jasak2007openfoam} is used in the simulations in this study. The flow is considered as viscous and incompressible. The governing equations are the two-dimensional Navier-Stokes equation and a non-dimensional continuity equation can be expressed as follows:
\begin{gather}
\frac{\partial u}{\partial t}+u\cdot(\nabla u)=-\nabla p+Re^{-1}\Delta u \label{eq:1},\\ 
\nabla\cdot u=0 \label{eq:2},
\end{gather}
where $\boldsymbol{u}$, $t$ and $p$ are the non-dimensional velocity, time, and pressure respectively, the characteristic length is the length of the square cylinder $D$, $\nu$ is the kinematic viscosity of the fluid and $\overline{U} = 2m/s$ is the mean velocity at the inlet. The Reynolds number is defined as ${Re} =\overline{U}D/\nu = 1000$.

The computational domain for the square cylinder is depicted in Fig. \ref{fig:2d_flow_over_square}. The inlet boundary ($\Gamma_{in}$) is characterized by a uniform velocity inlet boundary condition. Additionally, the outlet boundary ($\Gamma_{out}$) is specified as a pressure outlet, maintaining zero velocity gradient and a constant pressure. The no-slip constraint ($\Gamma_{W}$) is enforced on both the top and bottom surfaces of the domain and on the surface of the 2D square cylinder. The origin of the flow field coordinate system is established at the center of the square cylinder, which has a side width of $D$. The distance from the inlet to the center of the square cylinder is $10D$, while the outlet boundary is positioned at a distance of $30D$ from the center, ensuring the full development of the wake.

Drawing inspiration from \citet{wang2023dynamic}, we position probes strategically on the surface of the square cylinder. The objective of this placement is to gather pressure information from 24 designated probes on the surface used in the subsequent DRL training process. The accuracy of the grid has been validated as demonstrated in the work of \citet{yan2023stabilizing}.

\subsection{Numerical simulation configuration of 3D square cylinder}  \label{sec:cfd2}

\begin{figure}
\centering
\begin{overpic}[width=0.97\textwidth]{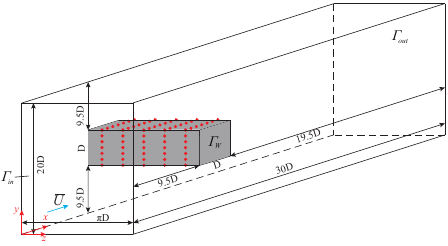}
\end{overpic}
\caption{Schematics of the computational domain for 3D turbulent flow around the square cylinder. The red points represent the locations of 120 surface pressure probes.} \label{fig:3d_flow_over_square}
\end{figure}

% In this section, four jets are placed at the corners of the square cylinder as actuators for AFC. The objective is to reduce the drag and lift forces of the square cylinder. 

%计算域
% 计算域的尺寸是根据 \citet{navroseFreeVibrationsCylinder2013}的研究设定的。如图所示，从入口（上游）和出口（下游）到圆柱体表面的距离分别设置为 10D 和 25.5D。此外，顶面和底面到圆柱体表面的距离均为 9.5D。方柱的跨距为 ΠD，如图所示。方柱表面上的red dot为压力探针，设为五圈，每圈24个，目的是得到充分的流场信息。
The computational domain dimensions are adapted following the investigation by \citet{trias2015turbulent}. Distances from the inlet (upstream) and outlet (downstream) to the surface of the square cylinder are set to 9.5$D$ and 19.5$D$, respectively. Additionally, the distances from the top and bottom surfaces to the surface of the square cylinder are 9.5$D$. The span of the square cylinders is $\pi D$, as depicted in Fig. \ref{fig:3d_flow_over_square}. The red points on the surface of the square cylinder represent pressure probes arranged in five circles, each containing 24 probes, ensuring comprehensive information about the flow field, as shown in Fig. \ref{fig:3d_flow_over_square}. Besides, $\overline{U} = 6.5m/s$ is the mean velocity at the inlet $\Gamma_{in}$ and ${Re} = 22000$. 

% 流动模拟采用大涡模拟（LES）技术进行。并使用有限体积法求解三维纳维-斯托克斯方程。模拟选择了PISOFOAM 求解器，这是一种采用PISO算法的不可压缩湍流的瞬态求解器.Crank-Nicolson 方案和二阶中心差分方案分别用于离散时间和空间导数。采用壁面适应局部涡流（WALE）粘度模型（Nicoud Ducros，1999 年）来模拟亚网格尺度应力。方柱表面采用了无滑动边界条件。在计算域的两侧和顶部实施了自由滑动条件，并在入口边界，速度值为均匀自由流速$\overline{U}$。计算域出口处应力矢量设置为零。在其余边界上，速度的法向分量和应力矢量的切向分量在两个方向上的值均为零。
The flow simulation employs the large eddy simulation (LES) technique \citep{zhiyin2015large}, solving the 3D Navier-Stokes equations through the finite volume method. The pisoFoam solver, a transient solver for incompressible turbulence using the PISO algorithm \citep{barton1998comparison} is selected. Discrete time and spatial derivatives utilize the Crank-Nicolson scheme and the second-order central difference scheme, respectively. A wall-adapted local eddy viscosity model \citep{nicoud1999subgrid} is employed for simulating subgrid-scale stresses. The surface of the square cylinder is subject to a no-slip boundary condition ($\Gamma_W$), ensuring fluid adherence. Free slip conditions are applied to the sides and top of the computational domain. At the inlet boundary ($\Gamma_{in}$), a uniform free-flow velocity $\overline{U}$ is prescribed. The stress vector at the outlet boundary of the computational domain ($\Gamma_{out}$) is set to zero. Other boundaries enforce zero values for the normal component of velocity and the tangential component of the stress vector in both directions. Throughout the time marching solution process, the position, velocity, and boundary conditions of the square cylinder are updated at each nonlinear iteration to accurately 
capture evolving flow dynamics.

\begin{figure} \centering
    \hfill
    \begin{overpic}[width=1\textwidth]{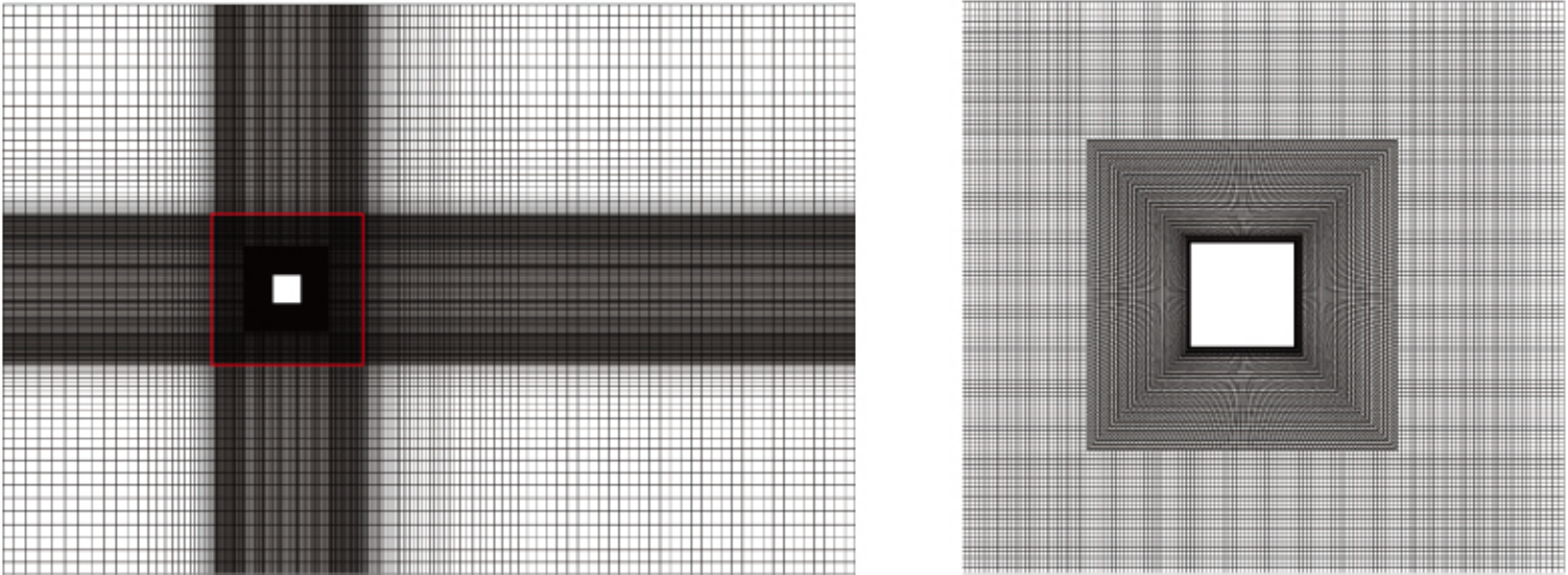}
        \put(-3,36){\fig{a}}
        \put(58,36){\fig{b}}
    \end{overpic}
    \hfill
\caption{Baseline grids used in the simulation: the full \fig{a} and partial \fig{b} computational domain.} 
\label{fig:mesh}
\end{figure}

%FIG 网格信息
This investigation explores turbulence surrounding a 3D square cylinder at $Re = 22000$. The computational domain is discretized using structured grids of varying sizes, illustrated in Fig. \ref{fig:mesh}. The grid is directly generated from the 3D simulation region to faithfully simulate physical phenomena and minimize numerical diffusion. Special attention is given to mesh refinement around the surface of the square cylinder and jets to enhance simulation accuracy. The study focuses on determining two non-dimensional aerodynamic coefficients: the drag coefficient ($C_D$) and lift coefficient ($C_L$):

\begin{equation}
C_D=\frac{2F_x}{\rho\overline{U}^2DH}, \quad C_L=\frac{2F_y}{\rho\overline{U}^2DH} \label{eq:cl},
\end{equation}
% 其中，Fx 和 Fy 分别为建筑物上的顺风（Fig. 中的 x 方向）和横风（图 1 中的 y 方向）空气动力，$\overline{U}$为入口平均风速。为验证网格的准确性，我们进行了网格调整研究。将三种不同网格设计的数值结果与不同仿真模拟数据进行了比较，网格收敛结果列于Table \ref{tab:3d_flow_validation}进一步考虑了与实际情况非常相似的更复杂的流动情况。列出了 Re= 22000 时模拟得出的 $C_D$、$C_L$ and $S_t$ 的统计值。三种网格在方柱周围采用了相同大小的计算单元，仅在尾流区域的单元大小有所不同。我们注意到，模拟结果与从文献中获得的值结果对网格分辨率并不敏感。
where $\rho$ is the density of the fluid in the flow field and $H = \pi D$ denotes the length of the square cylinder. $F_x$ and $F_y$ denote the aerodynamic forces acting in the downwind (x direction in Fig. \ref{fig:3d_flow_over_square}) and crosswind (y direction in Fig. \ref{fig:3d_flow_over_square}) directions on the square cylinder, respectively. To validate the accuracy of the grid, a grid verified study is conducted. Results from three distinct grid designs are compared with data from various simulations, and the grid convergence outcomes are summarized in Table \ref{tab:3d_flow_validation}. The statistical values of $C_D$, $C_L$, and $S_t$ derived from the simulation at $Re = 22000$ are presented. All three grids employ computational cells of the same size around the square cylinders, differing only in cell size in the wake region. It is noteworthy that the simulation results exhibit robustness to grid resolution, aligning with values obtained from existing literature \citep{trias2015turbulent,rodi1997status,voke1997flow}.

%表格：网格无关性
\begin{table}
    \centering
    \caption{Comparison of integral flow quantities in the flow past a 3D square cylinder at $Re = 22000$. $St=fD/ \overline{U}$ is the Strouhal number. LES$^i$ corresponds to each mesh resolution. Reference simulation results from literatures: DNS result\citep{trias2015turbulent};  LES results\citep{rodi1997status,voke1997flow};RANS results\citep{rodi1997status,voke1997flow}. } \label{tab:3d_flow_validation}
        \begin{tabularx}{\textwidth}{LCCCCC}
            \toprule
             & Mesh size & Mean $C_D$ & std of $C_D$ & std of $C_L$ & $S_t$ \\
            \midrule
            Present LES$^1$  & 1.4 million & 2.15 & 0.215 & 1.3 & 0.125 \\
            Present LES$^2$  & 2.8 million & 2.19 & 0.218 & 1.35 & 0.126 \\
            Present LES$^3$  & 5.2 million & 2.22 & 0.223 & 1.42 & 0.13 \\
            DNS result   & -   & 2.18 & 0.205 & 1.71 & 0.132 \\
            LES results   & -  & 2.02-2.77 & 0.14-0.27 &1.15-1.79& 0.09-0.15 \\
            RANS results\  & -  & 1.64-2.43 & $<0.27$ & 0.31-1.49 & 0.134-0.159 \\
            \bottomrule
        \end{tabularx}
\end{table}

\begin{figure} \centering
    \hfill
    \begin{overpic}[width=0.48\textwidth]{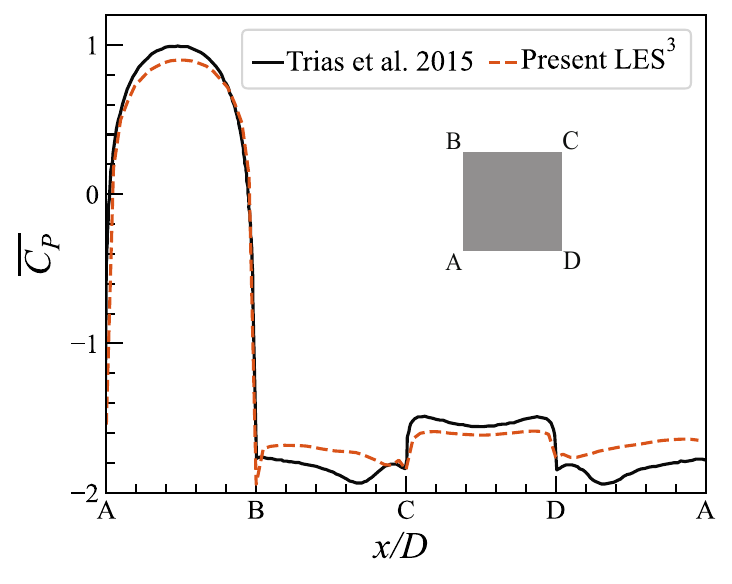}
        \put(0,75){\fig{a}}
    \end{overpic}
    \hfill
    \begin{overpic}[width=0.48\textwidth]{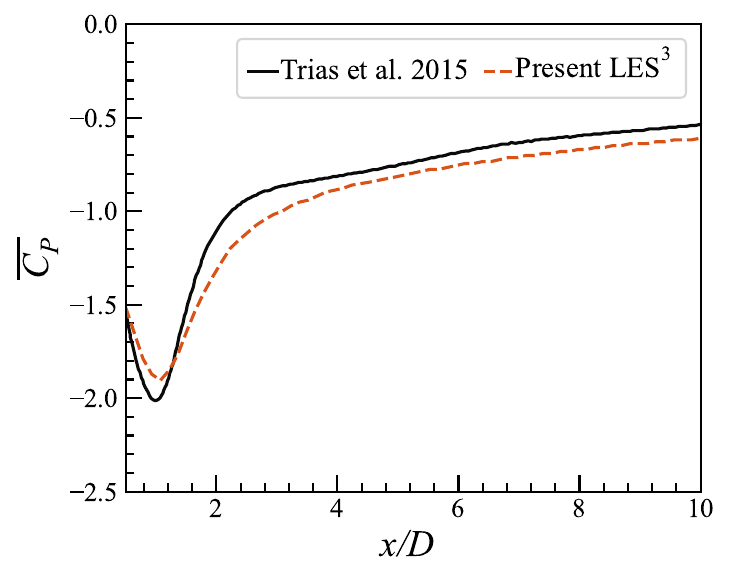}
        \put(0,75){\fig{b}}
    \end{overpic}
    \hfill
\caption{Comparison of mean pressure coefficient $C_P$ correspond to LES$^3$ result with DNS result of \citet{trias2015turbulent} corresponding to \fig{a} around the square cylinder and \fig{b} at domain centerline, respectively.} 
\label{fig:profile}
\end{figure}

% 验证profile
% 基线网格的 y+ 平均值保持在1以下，与 Saeedi Wang（2016 年）提出的建议一致。模拟中的最大库兰特-弗里德里希斯-路维数是动态的，保持在 0.15 以下，这确保了非稳态流在时间上得到解决。
% 为了进一步验证我们模拟的准确性，Fig. \ref{fig:profile}显示了方柱周围与尾流处压力系数$C_P$分布的平均值与DNS模拟研究的比较。可以看出，Present LES$^3$的数值结果在平均值方面与DNS模拟结果较吻合，同时在方柱尾流区也符合，只是略有出入。这进一步证实了我们模拟的可靠性。由于计算大网格仿真模拟所耗时间巨大，接下来的DRL训练中采用Present LES$^1$的网格，而使用Present LES$^3$进行测试，以节省计算成本。
For three mesh configurations, mean $y^+$ for the baseline mesh consistently remains below 1, aligning with the recommendation of \citet{saeedi2016large}. Throughout the simulation, the maximum Courant–Friedrichs–Lewy (CFL) number stays below 1, ensuring the temporal resolution of the unsteady flow. Fig. \ref{fig:profile} presents a comparison of the mean values of the pressure coefficient ($C_P$) distribution around the square cylinder and in the wake flow, contrasting the results with a DNS simulation study \citep{trias2015turbulent}. Notably, the numerical outcomes from Present LES$^3$ exhibit a favorable agreement with the DNS simulations concerning mean values, with minimal disparities in the surface and wake region of the square cylinder. This underscores the reliability of the simulations. Given the considerable time required for coarse grid simulations, the grid configuration of Present LES$^1$ is employed in the subsequent DRL training, while Present LES$^3$ is utilized for testing to optimize computational efficiency.

%FIG 四个jet的布置关系 3D
\begin{figure}
\centering
\begin{overpic}
[width=0.97\textwidth]{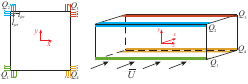}
\put(-5,30){\fig{a}}
\put(35,30){\fig{b}}
\end{overpic}
\caption{Details of four jet location and control configuration scheme in the \fig{a} 2D and \fig{b} 3D flow field of the square cylinder. The shaped regions on each corner side of the square cylinder represent the jet actuators $Q_1$, $Q_2$, $Q_3$, and $Q_4$. $l_{jet}$ represents the length of the jet actuator.} 
\label{fig:jets}
\end{figure}

\subsection{Configurations of flow control for the square cylinder}  \label{sec:cfd3}
%描述jet_location和设置
Considering the intricate and highly turbulent flow field around the square cylinder, the independence of each jet is set when implementing AFC. Illustrated in Fig. \ref{fig:jets}, four jet actuators $Q_{1}$, $Q_{2}$, $Q_{3}$, and $Q_{4}$ are strategically positioned at four corners of the square cylinder. Irrespective of the control mode (blowing or suction), the velocities of the four jets are consistently directed perpendicular to the surface of the square cylinder. Moreover, the DRL agent is deployed to generate four distinct actions for controlling each jet. This ensures that the velocity value at any corner of the square cylinder remains constant, aligning with the control objectives.

%描述jet长度
Based on uniformity in jet actuator control, the length of each jet consistently is maintained at $1/25$ of the corresponding side length. Regardless of the wind attack angle, the ratio $l_{jet}/D$ remains a constant of $1/25$. The mass flow rate $Q_i^*$ for each $Q_i$ is defined as:
\begin{equation}
Q_i^* = Q_{i}/ Q_{ref} =\frac{U_{jet,i}\cdot 2l_jet}{U \cdot D}  \label{flowrate}
\end{equation}
where $Q_i$ is the blowing and suction flow rate corresponding to Fig. \ref{fig:jets} and $U_{jet,i}$ is $i_{th}$ jet velocity. $Q_i^*$ is not greater than 0.037 in this study.

%action变化细节
% 为了避免使用 CFD 算法可能导致的压力和速度的非物理突变，控制机制采用了连续时间法。此外，在该系统中，选择适当的插值方法对于序列化接收到的时间离散化控制信号至关重要。因此，需要对喷射驱动进行平滑处理，以确保控制信号的连续变化，而不会因喷射速度的突然变化而产生过大的升力波动。每个动作步骤之间的射流速度变化过程是线性设定的：
To prevent non-physical abrupt alterations in pressure and velocity caused by jet actuator control, a continuous time approach is employed for the control mechanism in the CFD algorithms. Additionally, the proper choice of an interpolation method is pivotal for serializing the received time-discretized control signal in this system. Consequently, the jet actuation undergoes smoothing to guarantee a continuous transition in the control signal, preventing excessive lift fluctuations arising from sudden changes in jet velocity. The alteration process in jet velocity between each action step is linearly defined \citep{yan2023stabilizing}.
\begin{equation}
    V_{\Gamma_i(t)} = V_{\Gamma_i(t-1)}+ [a - a_{(t-1)}]/N_e,\quad i=1, 2,
\label{eq:action_process}
\end{equation}
where $V_{\Gamma_i(t)}$ is the jet flow velocity used at the non-dimensional times $t$, and $a$ is the jet velocity of $\Gamma_i$ for the current action time step.

\section{Reinforcement learning algorithms} \label{sec:DRL}
% 引用要加上，MIKT改写部分可以缩减，每句话设为不同的，加引用   more detail
% 在本节中，我们将首先介绍我整个框架和每个组件。然后，我们将介绍如何学习状态和动作嵌入，以及如何将多个跨域源策略自适应地转移到当前任务中。最后，我们将详细介绍SDTL与特定 DRL 算法 SAC [Schulman 等人，2017] 的结合。
In this section, the basic Soft Actor-Critic (SAC) algorithm \citep{haarnoja2018soft} is first introduced. Then, how to learn state embeddings and how to adaptively transfer one cross-domain source policy to learn the current policy is described. Finally, the state dimension mismatch transfer learning method (SDTL) combined with the basic SAC algorithm is illuminated in detail.

\subsection{Description of off-policy reinforcement learning} \label{sec:DRL1}
In this study, DRL within the context of an infinite-horizon discrete-time Markov Decision Process (MDP) which denoted as $\mathcal{M}=\langle\mathcal{S},\mathcal{A},\mathcal{R},{\mathcal{T}},\gamma\rangle$, where $\mathcal{S}$ and $\mathcal{A}$ represent the sets of states and actions, respectively. The state transition probability function is denoted as $\mathcal{T}:\mathcal{S}\times\mathcal{A}\times\mathcal{S}\mapsto[0,1]$, and the reward function, evaluating agent performance, is denoted as $\mathcal{R}:\mathcal{S}\times\mathcal{A}\times\mathcal{S}\mapsto\mathbb{R}$. The discount factor for future rewards is represented by $\gamma$. A policy $\pi:\mathcal{S}\times\mathcal{A}\mapsto[0,1]$ is defined as a state-conditioned probability distribution over actions, and the agent objective is to discover an optimal policy $\pi^{*}$ that maximizes the expected discounted return $R=\sum_{i=t}^{T}\gamma^{i-t}r_{i}$.

%介绍SAC  actorloss criticloss
% 熵（entropy）表示对一个随机变量的随机程度的度量。具体而言，如果是一个随机变量，且它的概率密度函数为，那么它的熵就被定义为$H(X)=\mathbb{E}_{x\sim p}[-\log p(x)]$. 在强化学习中，我们可以使用 来表示策略在状态下的随机程度。
The deep deterministic policy gradient (DDPG) off-policy algorithm \citep{lillicrap2015continuous} is extensively employed for estimating the gradient of the DRL objective. Soft Actor-Critic (SAC) \citep{haarnoja2018soft} is one of the most efficient off-policy algorithms and is designed for proficiently learning randomized policies and facilitating maximum entropy reinforcement learning. Entropy serves as a quantification of the randomness inherent in a random variable. Specifically, for a random variable $X$ with a probability density function denoted as $p$, its entropy $H$ is defined as $H(X)=\mathbb{E}_{x\sim p}[-\log p(x)]$. In the context of reinforcement learning, $H(\pi(\cdot|s))$ characterizes the extent of randomness in a policy within a given state.

% 最大熵强化学习（maximum entropyDRL）的思想就是除了要最大化累积奖励，还要使得策略更加随机。如此，强化学习的目标中就加入了一项熵的正则项，定义为
% 其中，是一个正则化的系数，用来控制熵的重要程度。熵正则化增加了强化学习算法的探索程度，越大，探索性就越强，有助于加速后续的策略学习，并减少策略陷入较差的局部最优的可能性。
The concept behind maximum entropy reinforcement learning involves not only maximizing cumulative rewards but also making the control strategy that exhibits increased randomness. To achieve this, an entropy regularization term is incorporated into the DRL objective, defined as:
\begin{equation}
\pi^*=\arg\max_\pi\mathbb{E}_\pi\left[\sum_tr(s_t,a_t)+\alpha H(\pi(\cdot|s_t))\right]
\label{eq:3.1}
\end{equation}
where $\alpha$ is a regularization factor to control the importance of entropy. Entropy regularization plays a pivotal role in enhancing the exploration aspect of the DRL algorithm. A higher value for $\alpha$ amplifies the exploratory nature, expediting subsequent policy learning and mitigating the risk of the policy converging to a suboptimal local minimum.

% 在最大熵强化学习框架中，由于目标函数发生了变化，其他的一些定义也有相应的变化。首先， Soft 贝尔曼方程如下：
Within the framework of Maximum Entropy Reinforcement Learning, the Soft Bellman equation is expressed as follows:
\begin{equation}
Q(s_t,a_t)=r(s_t,a_t)+\gamma\mathbb{E}_{s_{t+1}}[V(s_{t+1})]
\label{eq:3.2}
\end{equation}

% 其中，状态价值函数被写为
where the state value function is written as
\begin{equation}
V(s_t)=\mathbb{E}_{a_t\sim\pi}[Q(s_t,a_t)-\alpha\log\pi(a_t|s_t)]
\label{eq:3.3}
\end{equation}

\begin{figure}
\centerline{\includegraphics[width=0.6\textwidth]{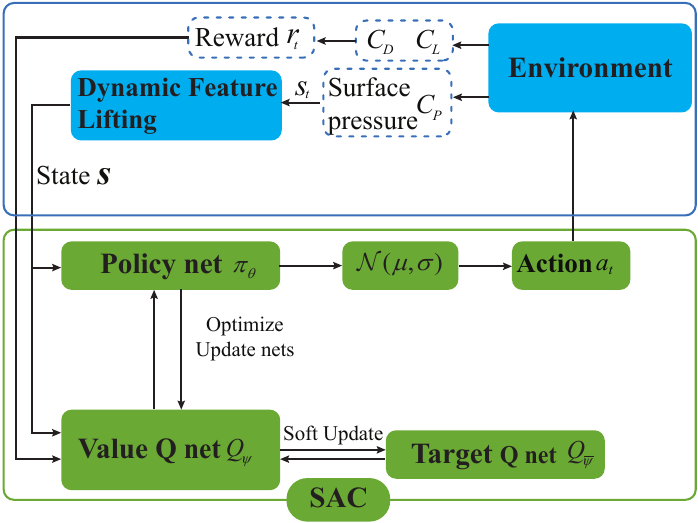}}
\caption{Schematic of SAC algorithm framework used in the present study. 
The Q network update: the target Q net $\mathcal{Q}_{\overline{\psi}}$ gets the target Q value based on $\mathbf{s}$, then the value Q net $\mathcal{Q}_{\psi}$ gets the desired Q value. The policy net $\pi_{\theta}$ update: Use the value Q net updated to calculate the desired value based on $\mathbf{s}$ and $r_t$. Then calculate the loss for the network parameter update and outputs $\mu$ and $\sigma$, which are used to sample action $a_{t}$. Noteworthily, the Dynamic Feature Lifting method is used from \citet{wang2023dynamic} in order to develop an effective flow state estimation that enables the DRL agent can obtain efficient information based on it. This framework is derived from the DRLinFluids package \citep{wang2022drlinfluids}.}
\label{fig:sac}
\end{figure}

% 在 SAC 算法中，我们为两个动作价值函数（参数分别为和）和一个策略函数（参数为）建模。基于 Double DQN 的思想，SAC 使用两个网络，但每次用$Q$网络时会挑选一个$Q$值小的网络，从而缓解值过高估计的问题。任意一个函数的损失函数为：
The SAC algorithm incorporates two value functions $Q$ parameterized by $\psi_1$ and $\psi_2$ respectively, along with a policy function $\pi$ parameterized by $\theta$, as illustrated in Fig. \ref{fig:sac}. Drawing inspiration from the concept of DDPG, the SAC employs two distinct networks for $Q$ function. Notably, each time a $Q$ network is utilized, it selects the network associated with a smaller $Q$ value. This approach mitigates the issue of value overestimation. The loss function for each value function is expressed as follows:
\begin{equation}
\begin{aligned}
L_{Q}(\psi) =\mathbb{E}_{(s_{t},a_{t},r_{t},s_{t+1})\sim \mathcal{D}}\left\lfloor\frac{1}{2}\left(Q_{\psi}(s_{t},a_{t})-(r_{t}+\gamma V_{\overline{\psi}}(s_{t+1}))\right)^{2}\right\rfloor  
\end{aligned}
\label{eq:3.4}
\end{equation}

% 其中，是策略过去收集的数据，因为 SAC 是一种离线策略算法。为了让训练更加稳定，这里使用了目标网络，同样是两个目标网络，与两个网络一一对应。SAC 中目标网络的更新方式与 DDPG 中的更新方式一样。% 策略的损失函数由 KL 散度得到，化简后为：
where $ \mathcal{D}$ is the data collected in the historical training process since SAC is an off-policy algorithm. To enhance training stability, the present networks $Q_{\overline{\psi}}$ are employed. Specifically, two current networks $Q$ are utilized sequentially, aligning with the two corresponding $Q$ networks. The updating procedure for the current networks in SAC mirrors that of DDPG. The loss function for the strategy function $\pi$ is derived from the Kullback-Leibler (KL) divergence and is subsequently simplified to:
\begin{equation}
L_\pi(\theta)=\mathbb{E}_{s_t\sim \mathcal{D},a_t\sim\pi_\theta}[\alpha\log(\pi_\theta(a_t|s_t))-Q_\psi(s_t,a_t)]
\label{eq:3.5}
\end{equation}

%FIG SAC流程图

In addition, as illustrated in Fig. \ref{fig:sac}, the SAC algorithm proceeds through each step by selecting four actions $a_{t}$ based on the current policy, executing these actions, and observing the resulting environmental feedback, including rewards $r_{t+1}$ and the subsequent state $S_{t+1}$. The state $S_{t+1}$ is then constructed as a state representation ($S$), incorporating historical time series data through the application of the Dynamic Feature Lifting method \citep{wang2023dynamic}. This involves leveraging pressure data from surface probes on the square cylinder collected over the preceding 30 action time steps. Notably, the performance of the DRL agent is significantly enhanced by transforming sensor signals into dynamic features. Consequently, the agent can effectively utilize surface pressure data instead of relying on velocity information from the wake.

\subsection{Learning state correspondence} \label{sec:DRL2}
%介绍SDTL   介绍必要的内容即可
In this section, the aim is to distill knowledge from a pre-trained source policy to a current policy with the aim of enhancing the sample efficiency of the ongoing learning process, which is presented from \citet{wan2020mutual}. The problem is set as considering scenarios where the source and current policies operate within distinct MDPs. The MDP properties $(\mathcal{S},\mathcal{A},\mathcal{R},{\mathcal{T}},\gamma)$ could be different, but a fundamental structural commonality persists between source and current MDPs. This commonality exemplified by the transfer from a 2D flow fluid to a 3D flow fluid \citep{he2023policy}, allows for the adaptation of knowledge. For notational simplicity, $\mathcal{S}_{\text{source}}$ and $\mathcal{S}_{\text{current}}$ denote the state spaces of the source and current MDPs, respectively. Here assume the availability of a pre-trained source policy network within the source MDP and concentrate on extracting representations from this source policy network that prove beneficial for learning within the current MDP. Moreover, the challenge of knowledge transfer is arised when $\mathcal{S}_{\text{source}} \neq \mathcal{S}_{\text{current}}$. To address this mismatch, a learned embedding space parameterized by an encoder function $\phi(\cdot)$ which is defined as $\mathcal{S}_{\text{emb}}:=\{ \phi(s)\mid s\in\mathcal{S}_{\text{current}}, \phi(s)\in\mathcal{S}_{\text{source}} \}$. Data points from this embedding space serve to extract valuable information from the source policy network. To ensure compatibility, the dimension of the embedding space is enforced to match the state space in the source MDP ($|\mathcal{S}_{\text{emb}}|$ = $|\mathcal{S}_{\text{source}}|$). It is worth noting that any embedding vector $s \in \mathcal{S}_{\text{emb}}$ is not necessarily a valid input state in the source MDP.

%TASK-ALIGNED EMBEDDING SPACE
Hence, the encoder parameter ($\phi$) plays a pivotal role in ensuring that the generated embeddings align with the DRL objective. In order to transfer from a source pre-trained policy to a current policy within the current MDP in disparate state spaces, lateral connections are incorporated between the current and source networks which can enhance representations in the layers of the current network by incorporating valuable representations from the layers of the source network. Consequently, this facilitates the extraction of transferable knowledge from both the source policy and state-value networks. Additionally, the source policy and value network are represented by $\pi_\theta'$ and $V_\psi'$, respectively, where the parameters ($\theta'$, $\psi'$) remain fixed throughout training. In parallel, ($\theta$, $\psi$) denote the trainable parameters for the current policy and value networks. As a general assumption for simplicity of exposition, the source and current policy and value networks are presumed to share the same number of hidden layers ($N_{\pi}$ and $N_Q$).
% Let $N_{\pi}$ and $N_Q$ represent the number of hidden layers in the source (and current) policy and value network, respectively.

In the current MDP, the current policy observes a state $s_{\mathrm{current}}\in\mathcal{S}_{\mathrm{current}}$ and then fed it into the encoder to produce state embedding $\phi(s_{\mathrm{current}})\in\mathcal{S}_{\mathrm{emb}}$, which can be easily passed through the source networks to extract $\{ z_{\theta^{\prime}}^{i},z_{\psi^{\prime}}^{i},1\leq i \leq N_{\pi},1\leq i \leq N_{Q} \}$ representing the pre-activation outputs of the $i$-th hidden layers of the source policy and value networks. Consequently, to acquire the hidden layer outputs $\{\boldsymbol{z}_{\boldsymbol{\pi}_\theta}^i, z_{Q_{\psi}}^i \}$ at layer $i$ in the current networks, the following relationships hold:
% under condition of $\left|\mathcal{S}_{\mathrm{emb}}\right|=\left|\mathcal{S}_{\mathrm{source}}\right|$. To obtain the pre-activation representations in the current networks, the state $s_{\mathrm{current}}$ is input, and a weighted linear combination is performed with the corresponding outputs and pre-activations from the source networks. 

\begin{equation}
\begin{aligned}
z_{\pi_\theta}^i&=\sigma\left(p_\theta^iz_\theta^i+(1-p_\theta^i)z_{\theta^\prime}^i\right),\\
z_{Q_\psi}^i&=\sigma\left(p_{\psi}^iz_{\psi}^i+(1-p_{\psi}^i)z_{{\psi}^\prime}^i\right)
\end{aligned}
\label{eq:3.6}
\end{equation}

where $\sigma$ is the activation function. The learnable mixing weights $\begin{aligned}\{p_\theta^i,p_\psi^i\}\in[0,1]\end{aligned}$ serve as dynamic factors that evolve over time, governing the diminishing influence of source policy on the current policy – a higher value of $\begin{aligned}\{p_\theta^i,p_\psi^i\}\end{aligned}$ signifies a reduced impact. These coefficients are initially set to low values during the early stages of training to facilitate the provision of essential information for starting the learning process. However, at the end of the training process, the current policy becomes entirely independent of the source policy, facilitating swift deployment during test time. To foster this independence, the additional coupling loss terms are designed to drive $\{p_\theta^i,p_\psi^i\}$ towards 1 as the training progresses:

\begin{equation}
\begin{aligned}
L^\pi_{\mathrm{coupling}}=-\frac{1}{N_{\pi}}\sum_{i=1}^{N_{\pi}}\log\left(p_{\theta}^{i}\right),
L^{Q}_{\mathrm{coupling}}=-\frac{1}{N_{Q}}\sum_{i=1}^{N_{Q}}\log\left(p_{\psi}^{i}\right)
\end{aligned}
\label{eq:3.7}
\end{equation}

% ENRICHED EMBEDDINGS WITH MUTUAL INFORMATION MAXIMIZATION
In order to drive the update of encoder parameters ($\phi$), it is imperative for the input states and state embeddings to demonstrate a highly correlated, thereby enabling optimal guidance from the source policy. So maximizing the mutual information between states and embeddings is the crucial approach. This objective is articulated as follows, drawing inspiration from \citep{wan2020mutual}:

\begin{equation}
\begin{aligned}
\mathcal{I}(\boldsymbol{s};\boldsymbol{e})& =\mathcal{H}(\boldsymbol{s})-\mathcal{H}(\boldsymbol{s}|\phi(s))  \\
&=\mathcal{H}(\boldsymbol{s})+\mathbb{E}_{\boldsymbol{s},\boldsymbol{e}}[\log p(\boldsymbol{s}|\boldsymbol{e})] \\
&=\mathcal{H}(\boldsymbol{s})+\mathbb{E}_{\boldsymbol{s},\boldsymbol{e}}[\log q_{\omega}(\boldsymbol{s}|\boldsymbol{e})] 
\\&+\mathbb{E}_{\boldsymbol{e}}\big[D_{\mathrm{KL}}(p(\boldsymbol{s}|\boldsymbol{e})||q_{\omega}(\boldsymbol{s}|\boldsymbol{e}))\big] \\
&\geq\mathcal{H}(\boldsymbol{s})+\mathbb{E}_{\boldsymbol{s},\boldsymbol{e}}[\log q_{\omega}(\boldsymbol{s}|\boldsymbol{e})]
\end{aligned}
\label{eq:3.8}
\end{equation}

where $\mathcal{H}$ denotes the differential entropy. The optimization objective outlined above is recognized as a variational information maximization algorithm, where the variational distribution $q_\psi(\boldsymbol{s}|\boldsymbol{e})$ serves as an approximation to the true conditional distribution $p(\boldsymbol{s}|\boldsymbol{e})$. Consequently, the ultimate optimization objective can be expressed as:

\begin{equation}
\begin{aligned}
L^\pi_{\mathrm{MI}}(\phi,\omega) &=-\mathbb{E}_{s\sim\rho_{\pi_{\theta}}}[\log q_{\omega}(s_\mathrm{current}|\phi(s_\mathrm{current}))]   \\
L^Q_{\mathrm{MI}}(\phi,\omega) &=-\mathbb{E}_{s\sim\rho_{Q_{\psi}}}[\log q_{\omega}(s_\mathrm{current}|\phi(s_\mathrm{current}))]
\end{aligned}
\label{eq:3.9}
\end{equation}
where $\rho_{\pi_\theta}$ and $\rho_{Q_{\psi}}$ is the state distribution of the current policy and value network.

%FIG SDTL-SAC细节图
\begin{figure}
\centerline{\includegraphics[width=0.7\textwidth]{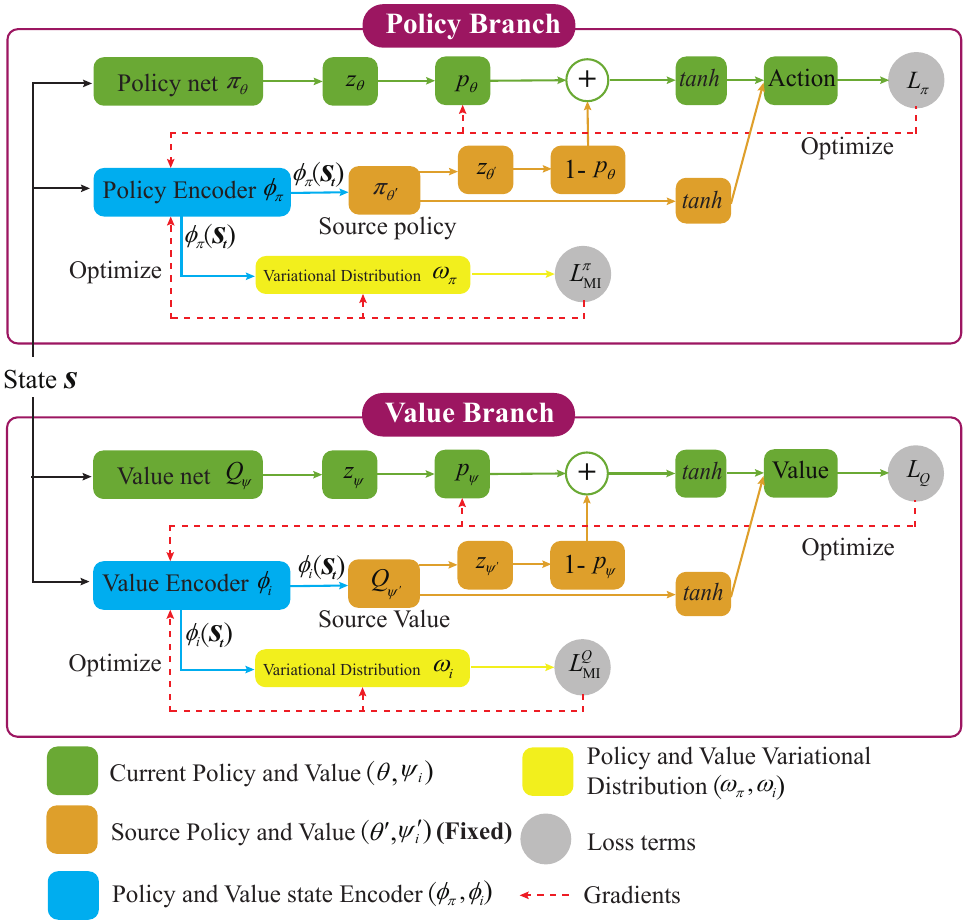}}
\caption{Schematic of the SDTL-SAC framework used in the present study. 
Within the current networks (green), pre-activation representations are blended linearly, utilizing learned mixing weights, with corresponding representations from the source policy, as detailed in Equations \ref{eq:3.6}. The source networks (orange) remain constant throughout training without undergoing gradient updates. The encoder parameters, denoted as $\phi_\pi$ ($\phi_i$) which depicted in blue, receive gradients from two distinct sources: the policy-gradient loss $L_\pi$ (or the value function loss $L_Q$) and the mutual information loss $L^\pi_{\mathrm{MI}}$ ($L^Q_{\mathrm{MI}}$). It is important to note that this knowledge flow occurs across all network layers.
}
\label{fig:sdttsac_workflow}
\end{figure}

\subsection{SDTL-SAC algorithm} \label{sec:DRL4}
%OVERALL ALGORITHM       
Fig. \ref{fig:sdttsac_workflow} illustrates the schematic diagram encompassing the comprehensive architecture and gradient flows, accompanied by a description of the implemented neural networks. This algorithm combines with Mutual Information based Knowledge transfer \citep{wan2020mutual,you2022cross} and basic SAC algorithm, specifically addresses state dimensional mismatch problems, named SDTL-SAC. In each iteration, the policy is executed within the current MDP, generating a batch of trajectories. This experience is then utilized to compute both the DRL loss (Equation \ref{eq:3.4} and \ref{eq:3.5}) and the mutual information loss (Equation \ref{eq:3.9}). The resulting gradients are employed to update various parameters. Simultaneously using these losses to update the encoder ($\phi$) ensures the fulfillment of the target for embeddings correlating them with the states in the current MDP. The coupling parameters $p_\pi$ and $p_\psi$, essential for the weighted combination of representations in the source and current networks, are updated using the coupling loss (Equation \ref{eq:3.7}) in conjunction with the DRL loss. The details of the SDTL-SAC algorithm are described in Appendix \ref{sec:SDTL-SAC}.

To illustrate the viability of employing cross-domain transfer learning in DRL for AFC, two distinct steps are undertaken as follows: first, the basic SAC algorithm is employed to train a control policy aimed at minimizing both $C_D$ and $C_L$; second, the SDTL-SAC algorithm is leverage for transfer learning, facilitating the transfer of control policy from a 2D square cylinder flow field to a 3D square cylinder flow field environment for the given application.

The study aims to reduce $C_D$ and $C_L$ of square cylinders by using DRL control. This objective can be achieved by setting an appropriate reward function. The reward function that combines $C_D$ and $C_L$ is proposed to achieve the optimization goal as follows:
\begin{equation}
    r_t =  - \langle C_D^{t} \rangle_T - |\langle C_L^{t} \rangle _T|, \label{eq:reward_function}
\end{equation}
where $C_D^{t}$ and $C_L^{t}$ means instantaneous $C_D$ and $C_L$ at time $t$ respectively, and $\langle \cdot \rangle _T$ indicates the mean value in one action period $T$ with DRL control.

\section{Results and discussion} \label{sec:Results}
% 这里补充一段话，加上两段的描述
The control performance and reliability of the basic SAC algorithm and a comparative analysis of the flow fields under baseline conditions and SAC control are shown in this section. Additionally, the process of transferring control policies when confronted with a state dimension mismatch from a 2D square cylinder to a 3D square cylinder is elucidated in Section \ref{sec:Results2}. Moreover, an investigation into the disparity in DRL training speed between SDTL-SAC and the basic SAC algorithm is conducted.

% \subsection{Analysis of the controlled flow}
\subsection{Flow control results using the basic SAC algorithm} \label{sec:Results1}
%FIG cd cl psd四张图
\begin{figure} \centering
    \begin{overpic}[width=0.48\textwidth]{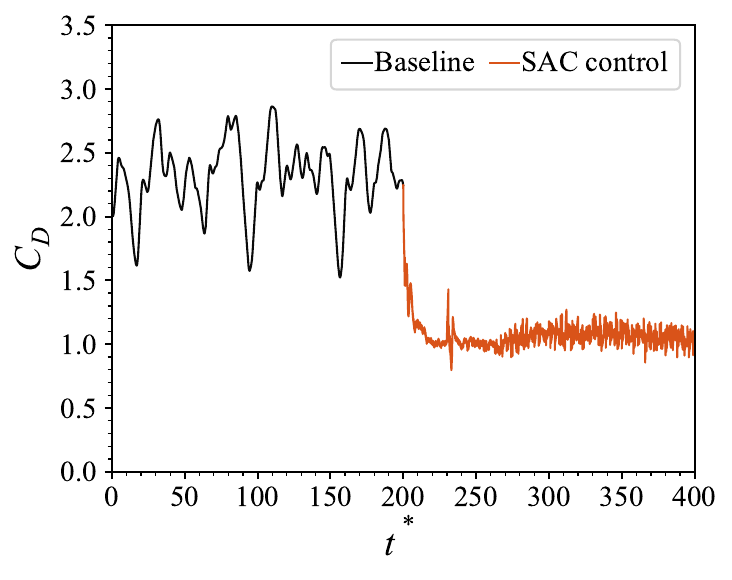}
        \put(0,75){\fig{a}}
    \end{overpic}
    \hfill
    \begin{overpic}[width=0.48\textwidth]{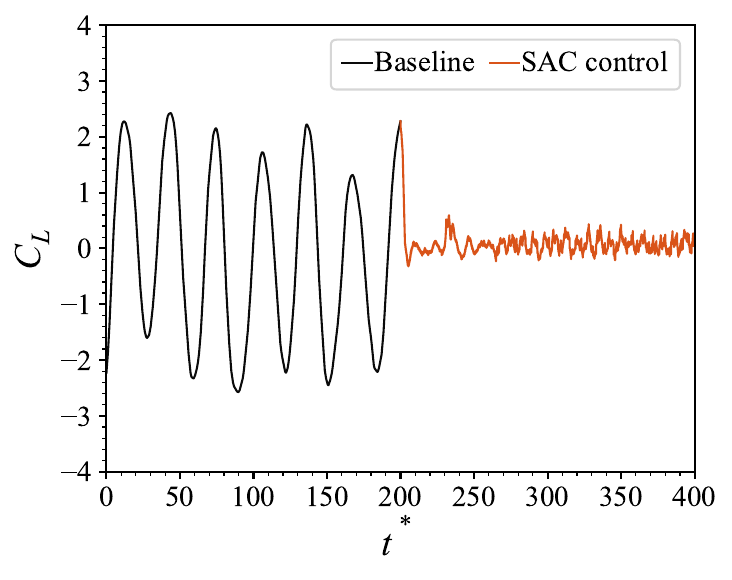}
        \put(0,75){\fig{b}}
    \end{overpic} \\
    \begin{overpic}[width=0.48\textwidth]{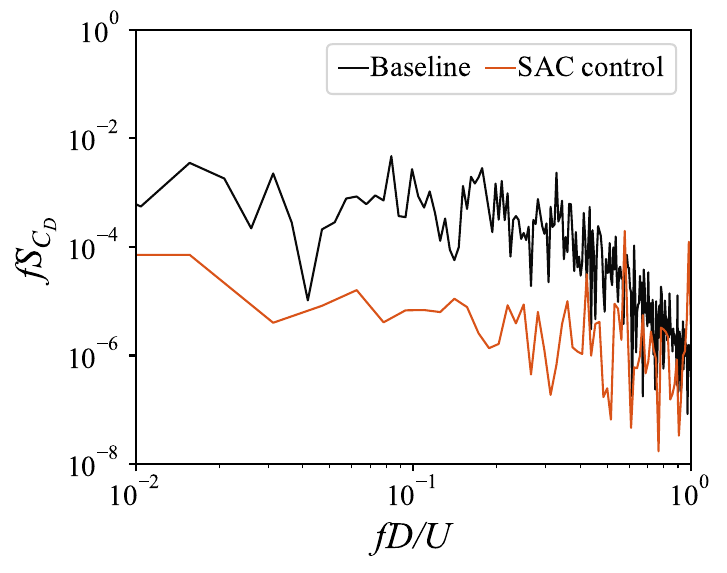}
        \put(0,75){\fig{c}}
    \end{overpic}
    \hfill
    \begin{overpic}[width=0.48\textwidth]{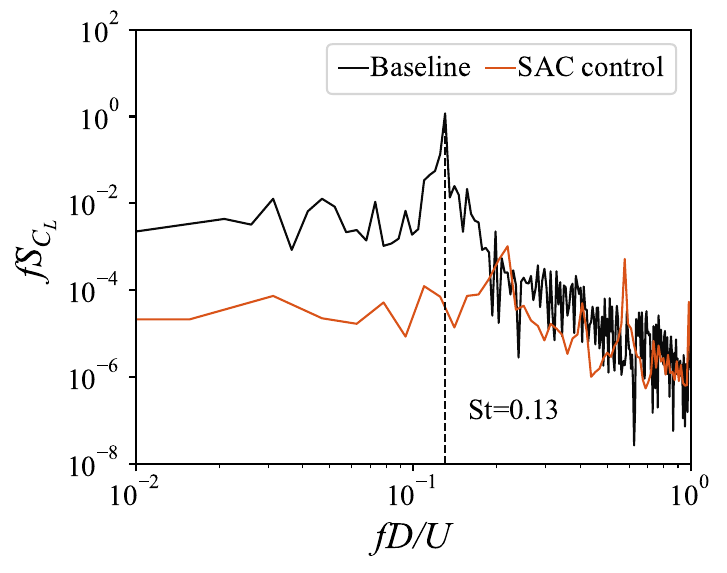}
        \put(0,75){\fig{d}}
    \end{overpic}
\caption{\fig{a} Evolution of $C_D$ for the 3D square cylinder without and with SAC control at $Re = 22000$; \fig{b} Temporal variations in smoothed $C_L$ for the square cylinder without and with SAC control at $Re = 22000$; \fig{c} and \fig{d} are the power spectral density (PSD) of $C_D$ and $C_L$ respectively during the period of non-dimensional time ranging from 300 to 400. $t^*$ represents non-dimensional time (one action step).} \label{fig:22000cdcl}
\end{figure}

%cdcl
% 在经过大量的DRL训练后，使用mutil jets control configuration得到收敛后的策略。为了全面评估代理的控制能力和鲁棒性，在测试阶段，代理与环境之间总共进行了 200 次交互。这种全面的评估对各种流场下的代理性能进行了彻底检查，进一步比较了它们在高雷诺数下控制方柱流场的有效性。
% 在经过充足的训练后,图{ref{fig:500_cdcl}展示了基于动态特征的SAC算法的结果。整个训练过程在 DRLinFluids 提供的10个环境中并行进行。该算法成功地学会了执行主动流控制，从而持续减少阻力并抑制升力。在没有驱动的情况下，$C_D$和$C_L$围绕一个平均值周期性地摆动，如图所示。$C_D$ 的平均值为 2.22，$C_D$与$C_L$ 的 std 值分别为 0.223， 1.42。采用基于 DF-DRL 的主动流量控制后，平均值 $C_D$ 降至1.06，相当于阻力减少了约 $52.3\%$。此外，$C_D$与$C_L$ 的波动几乎被完全抑制，其 std 值分别降低到了 0.058 and 0.118.相比baseline，降幅分别为74\%和91.7\%, 
% 功率谱分析比较了有主动流量控制和无主动流量控制气缸的 $C_D$ 和 $C_L$，结果如图所示。方柱的$C_L$频谱在强迫频率 St = 1.3处出现峰值.这表明在这个频率上有一系列明显的涡流脱落，产生了平均阻力和波动所需的大部分能量。相比之下，采用基于 DF-DRL 的主动流控制的气缸的 $C_D$ 和 $C_L$ 功率谱曲线中的峰值消失了。说明规则的涡流脱落模式被抑制。另外，控制后的$C_D$ 和 $C_L$低频区域小于baseline,这与$C_D$ 和 $C_L$波动的降低是一致的。说明基于DRL控制器主要通过抑制相对低频区域的能量来抑制阻力和升力。
Following extensive DRL training epochs, the converged policy is convergence by employing four jet control configurations. To comprehensively assess the control efficacy and robustness of the agent, a total of 200 interactions are conducted between the agent and the environment during the testing phase. This evaluation affords a detailed analysis of the agent performance across diverse flow conditions and facilitates a comparative study of their effectiveness in controlling the square cylinder flow field at high $Re$.

Fig. \ref{fig:22000cdcl} \fig{a} and \fig{b} present the outcomes of the SAC algorithm. The entire training procedure occurs concurrently in 10 environments facilitated by DRLinFluids \citep{wang2022drlinfluids}. The algorithm effectively learns to perform AFC, consistently diminishing drag and suppressing lift. In the baseline case, both $C_D$ and $C_L$ exhibit periodic oscillations around their mean values. Specifically, the mean value of $C_D$ is 2.22, with standard deviations of 0.223 for $C_D$ and 1.42 for $C_L$. With the DRL-based control, the mean value of $C_D$ diminishes to 1.06, corresponding to a remarkable drag reduction of approximately $52.3\%$. Furthermore, the fluctuations in $C_D$ and $C_L$ experience significant suppression, with standard deviations reduced to 0.058 and 0.118, respectively. This represents a reduction of 74\% and 91.7\% compared to the baseline, respectively.

The power spectral density (PSD) analysis, comparing the behavior of $C_D$ and $C_L$ in cylinders with and without control, is presented in Fig. \ref{fig:22000cdcl} \fig{c} and \fig{d}. The $C_L$ spectrum exhibits a peak at the dominant frequency $St = 1.3$, indicating distinct vortex shedding at this frequency, which contributes significantly to the energy responsible for both mean drag and fluctuations. In contrast, the PSD curves for $C_D$ and $C_L$ in the cylinder under SAC control show a disappearance of peaks, signifying the suppression of the regular vortex shedding pattern. Furthermore, the low-frequency regions of $C_D$ and $C_L$ are smaller than the baseline, aligning with the reduction in the fluctuations of $C_D$ and $C_L$. This suggests that DRL-based control primarily diminishes drag and lift by suppressing energy in the relatively low-frequency range.

%与其他减阻效果进行比较

%FIG mean Cp  rmsCp  五圈都画出来看看   以及中间表面压力变化
\begin{figure} \centering
    \begin{overpic}[width=0.48\textwidth]{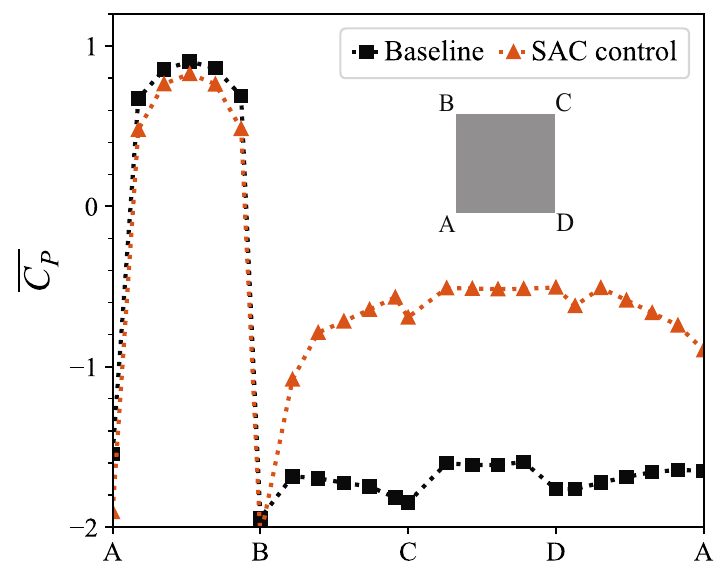}
        \put(2,75){\fig{a}}
    \end{overpic}
    \hfill
    \begin{overpic}[width=0.48\textwidth]{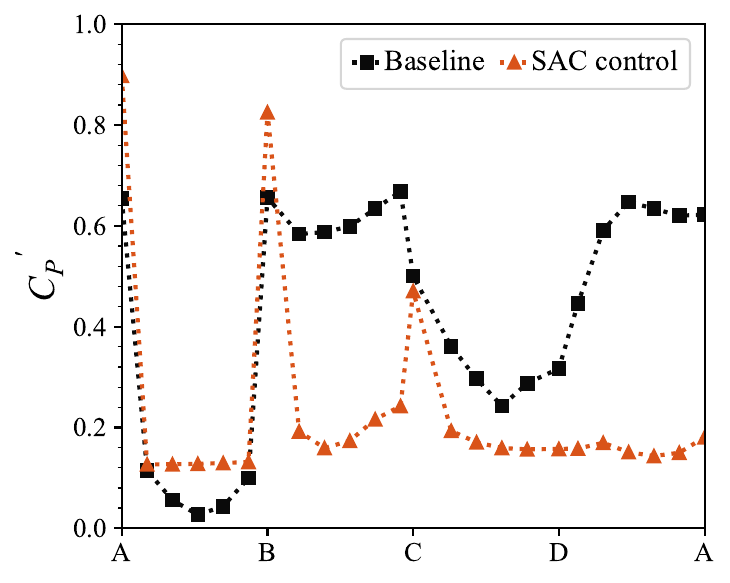}
        \put(1,75){\fig{b}}
    \end{overpic} \\
    % \begin{overpic}[width=0.48\textwidth]{Figures/psd_sidepressure.pdf}
    %     \put(2,73){\fig{c}}
    % \end{overpic}
\caption{Mean and fluctuating pressure coefficient $C_P$ distributions around the square cylinder for \fig{a} the baseline case and \fig{b} SAC control case. 
% \fig{c} depicts PSD of the square cylinder side-force fluctuation $C_P$
} 
\label{fig:22000pressure}
\end{figure}

% 如图（ref{fig:surface_pressure}）所示，方柱表面的平均压力系数（$C_P$）和波动压力系数（$C_P$）分别为基线情况和DRL AFC情况。在迎风面上，Baseline和DRL控制情况下的平均 $C_P$ 差异都很小。这种相似性意味着喷射流对方柱迎风面风压的影响可以忽略不计。然而，在其他三个面，尤其是背风面，观察到了明显的差异。在这些区域，与基线情况相比，AFC情况下的负 C_P$ 要小得多。此外，由于控制后背风面的负$C_P$降低，then $C_D$也降低。该面负压的减少有助于形成更有利的压力分布，从而改善流动控制并减少方柱上的阻力。
% 波动的 $C_P$ 也显示在图（ref{fig:surface_pressure}）中。在DRL控制情况下，迎风面上波动的 $C_P$的波动降低了一些。特别是对于背风面，与基线情况相比，AFC 情况明显抑制了 $C_P$ 的波动。这也解释了$C_L$的std显著下降。值得注意的是，AFC情况下在方柱三个角的位置显示出$C_P$ 波动的峰值。这些峰值是由射流的吹吸效应引起的涡流湍动造成的。尽管如此，这些峰值并不影响波动 $C_P$ 的整体下降，因为使用这种多射流配置控制仍然有效地降低了波动 $C_P$。
Fig. \ref{fig:22000pressure} shows the mean and fluctuating $C_P$ on the surface of the square cylinder in both the baseline and the SAC control cases. The mean $C_P$ exhibits minimal variation on the windward surface, suggesting that the jet flow has a negligible impact on the wind pressure in this region. However, notable differences emerge on the other three sides, particularly the leeward side. Besides, the negative $C_P$ is significantly smaller in the control case compared to the baseline, leading to a reduction in $C_D$. The decrease in negative pressure on the leeward face enhances the overall pressure distribution, contributing to improved flow control and reduced drag on the square cylinder.

The fluctuations of $C_P$ are depicted in Fig. \ref{fig:22000pressure} \fig{b}. On the windward side, the SAC control case exhibits a modest increasing in $C_P$ fluctuations. Notably, on the leeward surface, the control case substantially suppresses $C_P$ fluctuations compared to the baseline case, providing an explanation for the significant decrease in the standard deviation of $C_L$. It is noteworthy that the control case shows peaks in $C_P$ fluctuations at the three corners of the square cylinder, attributable to vortex turbulence induced by the jet blowing and suction effect. However, these peaks do not compromise the overall reduction in $C_P$ fluctuations, underscoring the effectiveness of the multiple jet configuration in controlling and minimizing $C_P$ fluctuations.

%FIG 控制前后的瞬时涡量图2D中间位置   控制前后瞬时涡量图
\begin{figure} \centering
    \begin{overpic}[width=1\textwidth]{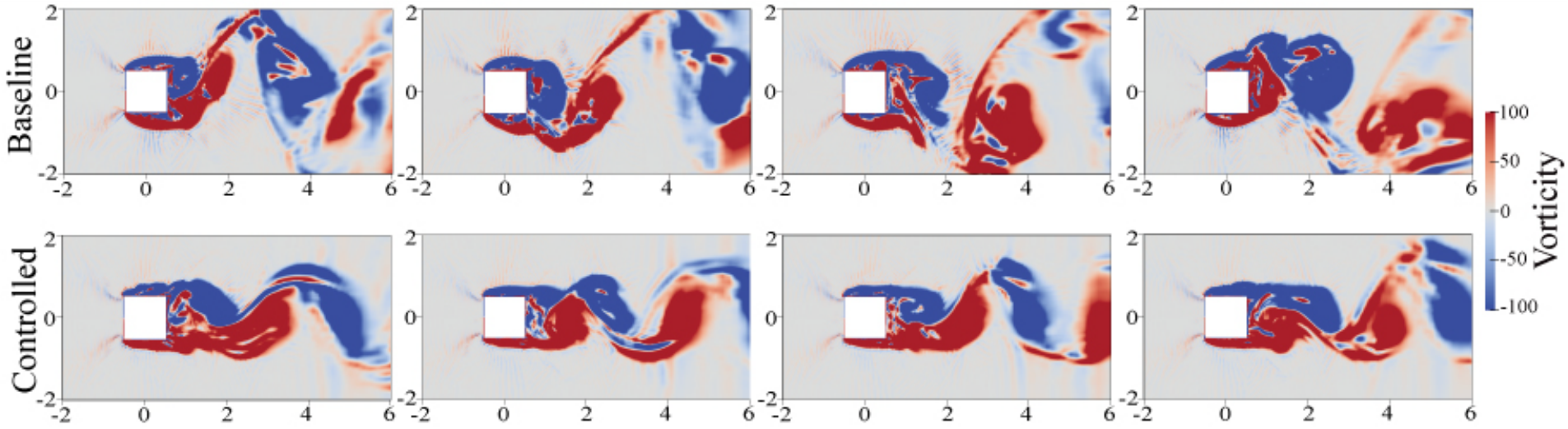}
    \put(-3,28){\fig{a}}
    \put(-3,13){\fig{b}}
    \end{overpic}
\caption{Instantaneous snapshots vortical structures close to the surface of the square cylinder colored by the streamwise vorticity \fig{a} without and \fig{b} with SAC control} 
\label{fig:shunshi}
\end{figure}

%FIG 分析四个jet的action与流量   %FIG 分析单纯四个角单纯吹吸参数实验情况
\begin{figure} \centering
    \begin{overpic}[width=0.8\textwidth]{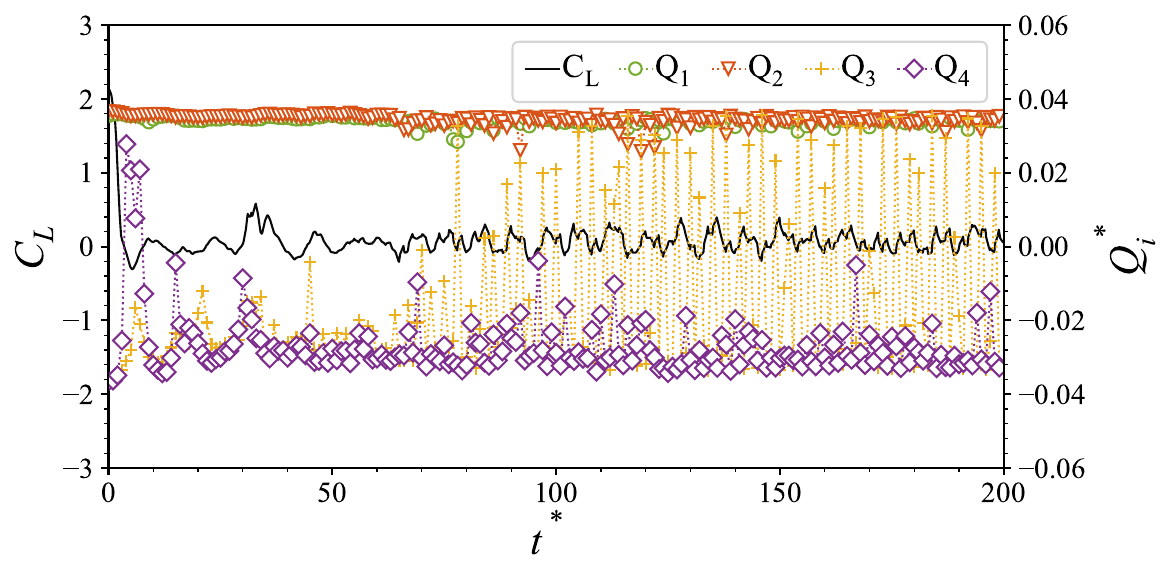}
    \end{overpic}
    \hfill
    \caption{Time histories of $C_L$, and the mass flow rate corresponding to each jet actuator during SAC control period.} \label{fig:actioncl}
\end{figure}

%控制前后的瞬时涡量   
% 图（ref{fig:shunshi}）描述了在200$t^*$ 内的 $t = T_s/4$、$T_s/2$、$3T_s/4$ 和 $T_s$ 时刻的瞬时涡度。
% 无控制下的流场小涡容易卷曲形成大涡，并往后传导，形成交替脱落的模式。且在方柱后缘处，形成的涡流通道较长
% 该吹吸气设置采用了利用后缘喷流的控制策略。能有效地利用前缘喷流（$Q_1$ 和 $Q_2$）分解涡流，将大涡先被破碎成小涡，随后利用后缘喷流（$Q_3$ 和 $Q_4$）将较小的涡流推离背风尾流区域，此时小涡可以附着在矩形柱表面不再卷曲形成大涡，从而形成更有规律的同步脱落涡。也更稳定。整体情况为小涡依次附着在方柱表面，沿流向形成稳定的涡流通道，同时形成对称脱落的尾流，其中流场形成的涡脱轨迹沿中心线对称. 因此方柱升阻力都能够大幅减少。
%jets速度分析
% 此外，图（ref{fig:actioncl}）显示了control 情况下 DRL代理在200$t^*$ 期间输出的 $C_L$ 和喷流作用。方柱四个角上的吹吸气质量流量变化规律有所不同。具体来说，$Q_1$ 和 $Q_2$ 完全处于吸气状态，而 $Q_3$ 和 $Q_4$ 呈现出波动的吹吸气状态。从$C_L$变化趋势可以看出，在80$t^*$后$C_L$表现为伪周期的状态。而此时$Q_4$的波动较为剧烈，吹气与吸气交替进行。说明后缘吹吸气的波动变化有利于$C_L$的稳定。且$C_L$ 的波动几乎完全被抑制，这突出表明了该策略在稳定涡流通道方面的稳健性。这强调了 DRL 控制的精确性。
Fig. \ref{fig:shunshi} illustrates the instantaneous vorticity moments with and without control. In the uncontrolled flow field, small vortices tend to coil up, forming larger vortices that propagate backward, resulting in an alternating shedding pattern. Notably, at the trailing edge of the square cylinder, the vortex channels exhibit increased length. The blowing and suction setup implements a control strategy utilizing the trailing-edge jet. This strategy efficiently utilizes the leading edge jets ($Q_1$ and $Q_2$) to disrupt the vortices, initially breaking down large vortices into smaller ones. Subsequently, the trailing edge jets ($Q_3$ and $Q_4$) are employed to push the smaller vortices away from the leeward wake region. This prevents the small vortices from attaching to the surface of the rectangular cylinder and forming large vortices, resulting in a more regular and synchronized shedding of vortices that is also more stable. The sequential attachment of small vortices to the surface of the square cylinder, forming a stable vortex channel along the flow direction, leads to the symmetrical shedding of vortices in the wake. As a result, the flow field formed by the shedding vortices follows a trajectory along the centerline symmetry, significantly reducing the drag and lift force of the square cylinder.

Furthermore, Fig. \ref{fig:actioncl} illustrates $C_L$ and jet actions of the DRL agent over a span of 200$t^*$ for the control case. The blowing and suction mass flow rates exhibit varying patterns at the four corners of the square cylinder. Specifically, $Q_1$ and $Q_2$ consistently display inspiratory behavior, while $Q_3$ and $Q_4$ exhibit fluctuating states of blowing and suction. Examining the trend of $C_L$ reveals a pseudo-periodic state emerging after 80$t^*$. Notably, the fluctuation of $Q_4$ becomes more pronounced during this period, displaying alternating blowing and suction. This observation suggests that the fluctuating dynamics of blowing and suction at the trailing edge contribute to the stabilization of $C_L$. The fluctuation of $C_L$ is effectively suppressed, underscoring the robustness of the strategy in stabilizing the vortex channel. This emphasizes the precision and effectiveness of the SAC control.

%FIG 控制前后的3D涡量图  流线图 
\begin{figure} \centering
    \hfill
    \begin{overpic}[width=0.45\textwidth]{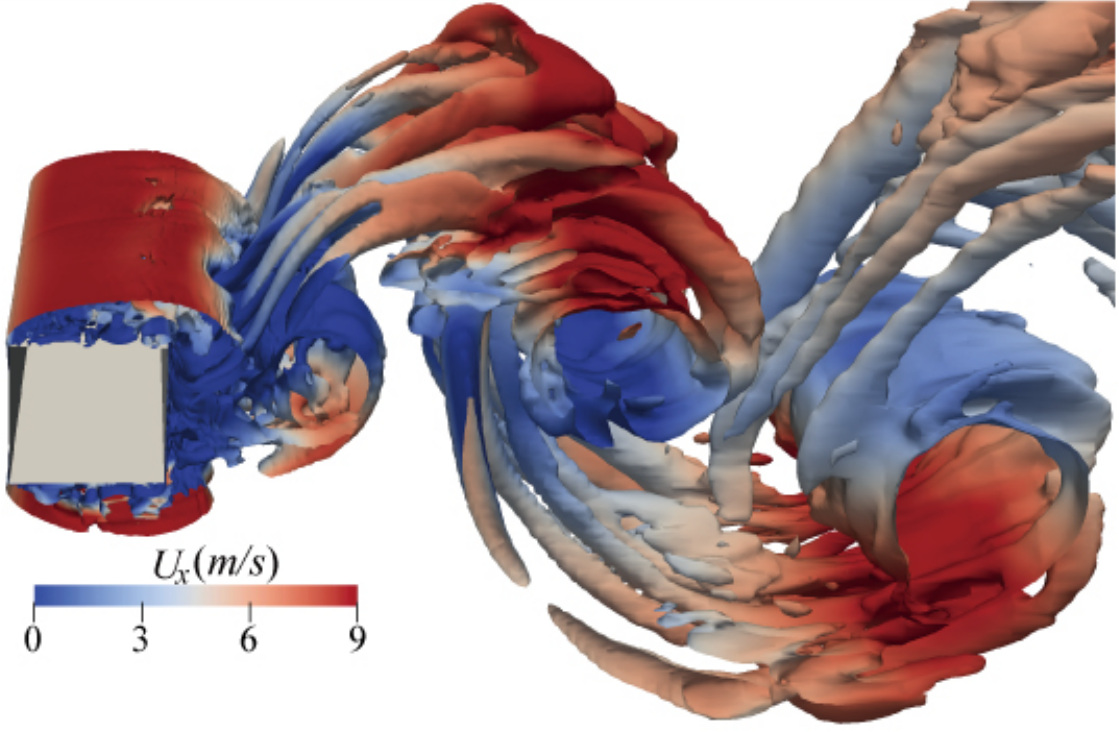}
        \put(-5,50){\fig{a}}
    \end{overpic}
    \hfill
    \begin{overpic}[width=0.45\textwidth]{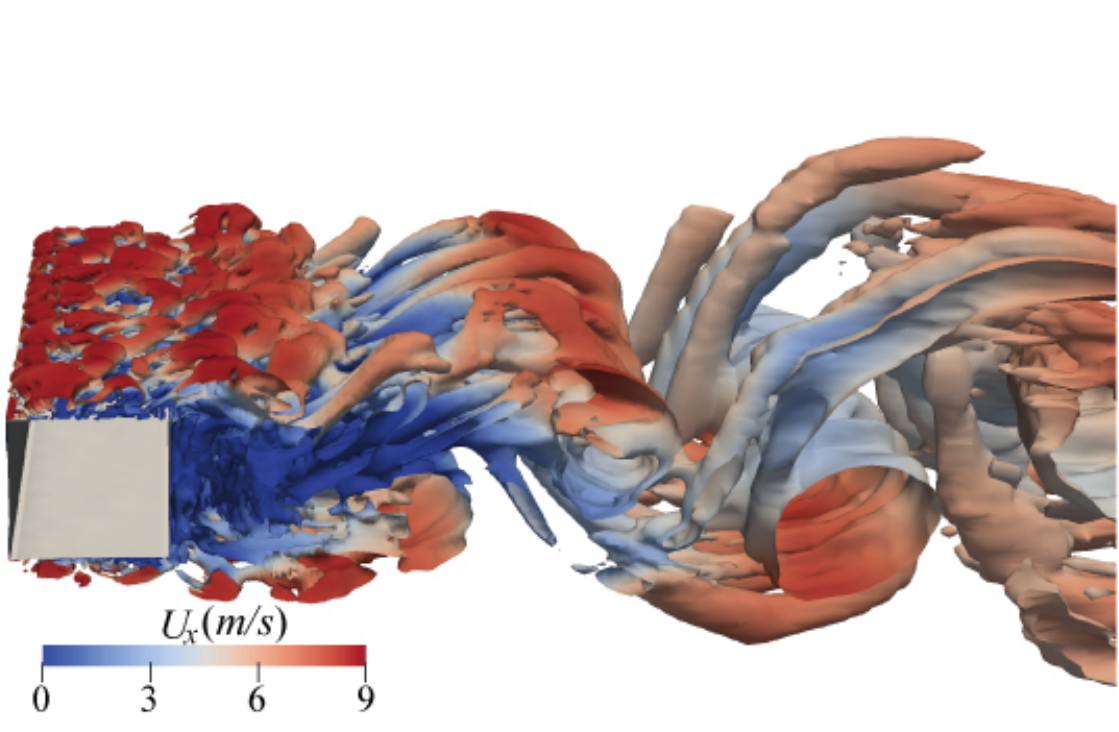}
        \put(-9,50){\fig{b}}
    \end{overpic}
    \hfill
    \\ \vspace{1em}
    \hfill
    \begin{overpic}[width=0.44\textwidth]{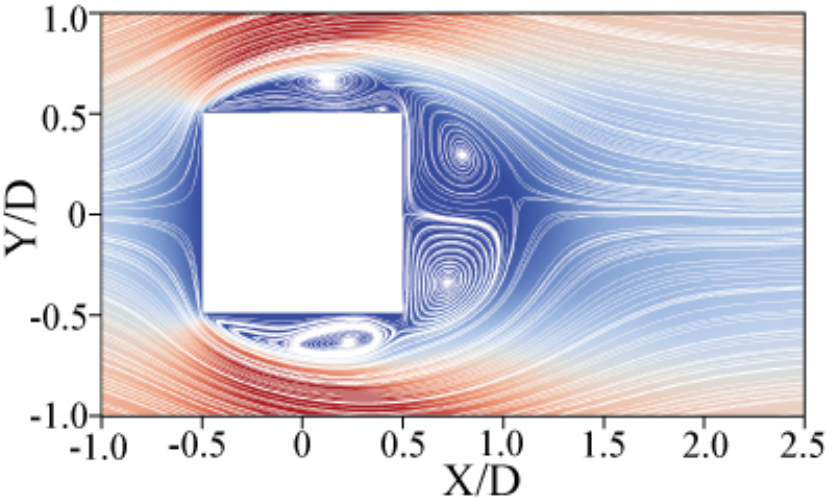}
        \put(0,64){\fig{c}}
    \end{overpic}
    \hfill
    \begin{overpic}[width=0.5\textwidth]{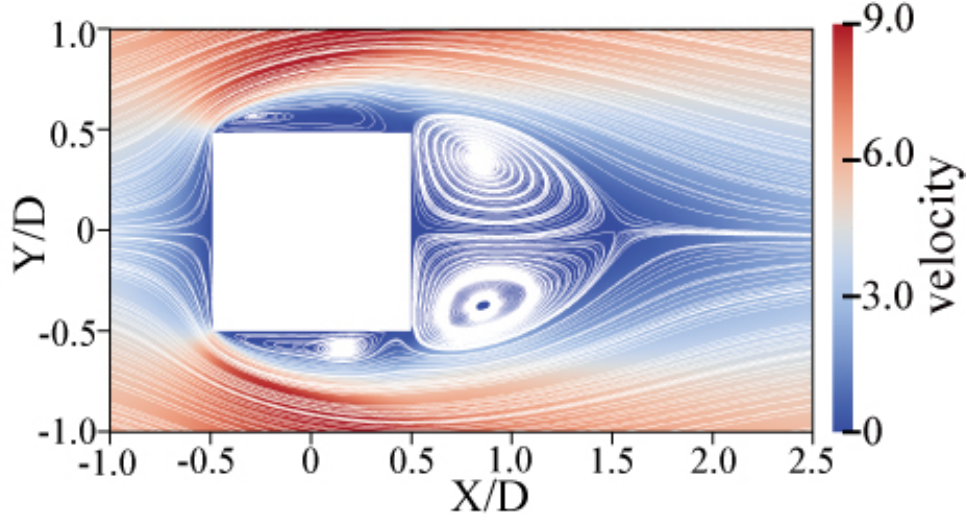}
        \put(2,55){\fig{d}}
    \end{overpic}
    \hfill
\caption{Turbulent flow past a square cylinder at Re = 22000. Instantaneous snapshots for iso-surfaces of Q-criterion colored by the streamwise velocity \fig{a} without and \fig{b} with SAC control, and streamlines of mean flow \fig{c} without and \fig{d} with SAC control.} 
\label{fig:q_and_mean_streamline}
\end{figure}

%Q值图    需要更改
% 从尾流场三维涡旋结构的角度来看，如图（ref{fig:q_and_mean_streamline}\fig{a}）和（fig{b}）所示，在没有喷流喷射的情况下，靠近方柱的尾流区域显然会产生细小而零散的涡旋。随着气流远离圆筒，小尺度涡旋结构逐渐消散，为大尺度涡旋让路。然而，当采用基于DRL的致动器进行控制时，由于方柱四角的吹吸作用，再加上流入的气流，会形成一个更规则的细长漩涡结构。
% 这种拉长的漩涡结构类似于致动器产生的条状结构，在一定程度上起到了减少碎裂漩涡产生的作用。因此，中远尾流区域呈现出更有规律的交替拉长漩涡模式。另外，可以明显的看出前缘吹吸气的影响，导致大尺度的流向辫状涡流大部分被小尺度的细碎涡流所取代，而后缘吹吸气将小涡流推离，形成的尾流变得更窄，这是CD变小的迹象。使用基于DRL的致动器引入了受控气流，对尾流特性产生了重大影响。拉长涡旋结构的形成更有规律性和连贯性，表明气流控制能力得到了提高。这种受控涡流结构有可能在各种工程应用中提高空气动力性能、减少阻力或实现其他所需的流动控制目标。
% 尾流再循环区的尺寸，特别是宽度和长度，直接影响方柱后面的基础压力。如图（ref{fig:q_and_mean_streamline}/fig{c}和（fig{d}）所示，由基于DRL的致动器控制的气缸尾流区域与未受控气缸相比具有以下特征：（1）次级涡缩小；（2）尾流最大宽度减少11.8%；（3）尾流长度延长80%。三个因素均导致$C_D$显著下降，且次级涡的缩小降低了对$C_L$的不利影响。这种总体效应表现为阻力的减少，导致阻力减少了约 52.3%。
Considering the 3D vortex structure in the wake field, depicted in Figs. \ref{fig:q_and_mean_streamline} \fig{a} and \fig{b}, it becomes evident that in the absence of jet control, the wake region proximate to the square cylinder generates small and dispersed vortices. As the inlet flow progresses away from the cylinder, the small-scale vortex structures gradually dissipate, giving way to larger scale vortices. Conversely, when the DRL-based jet control is employed, a more regular and elongated vortex structure emerges. This result stems from the blowing and suction actions at the corners of the square cylinder, working in tandem with the inflowing airflow.

The elongated vortex structure which is reminiscent of the strip structure generated by the actuator, effectively mitigates the generation of fragmented vortices. Consequently, the mid and far wake regions exhibit a more orderly pattern characterized by alternating elongated vortices. Notably, the influence of leading edge blowing and suction is evident which can transform large-scale flow braid vortices into predominantly smaller and finely fragmented vortices. Concurrently, trailing edge blowing and suction displace these small vortices, forming a narrower wake, indicative of reduced drag. The integration of a DRL-based jet actuator introduces controlled airflow that significantly shapes wake characteristics. The resulting elongated vortex structure exhibits enhanced regularity and coherence, pointing towards improved airflow control. These controlled vortex structures hold the potential to enhance aerodynamic performance, minimize drag, or achieve other desired flow control objectives across diverse engineering applications.

The width and length of the wake recirculation zone exert a direct influence on the base pressure behind the square cylinder. Illustrated in Fig. \ref{fig:q_and_mean_streamline} \fig{c} and \fig{d}, the wake region of the cylinder subject to DRL-based actuation exhibits (1) a constriction of the secondary vortex, (2) an 11.8\% reduction in the maximum width of the wake stream, and (3) an elongation of the wake stream length by 80\% compared to the uncontrolled cylinder. These three factors collectively contribute to a significant decrease in drag, and the constriction of the secondary vortex mitigates the adverse impact on $C_L$. This holistic effect results in a pronounced drag reduction of approximately 52.3\%.

%FIG 控制前后的pod分析   3Dpod  
\begin{figure} \centering
    \hfill
    \begin{overpic}[width=1\textwidth]{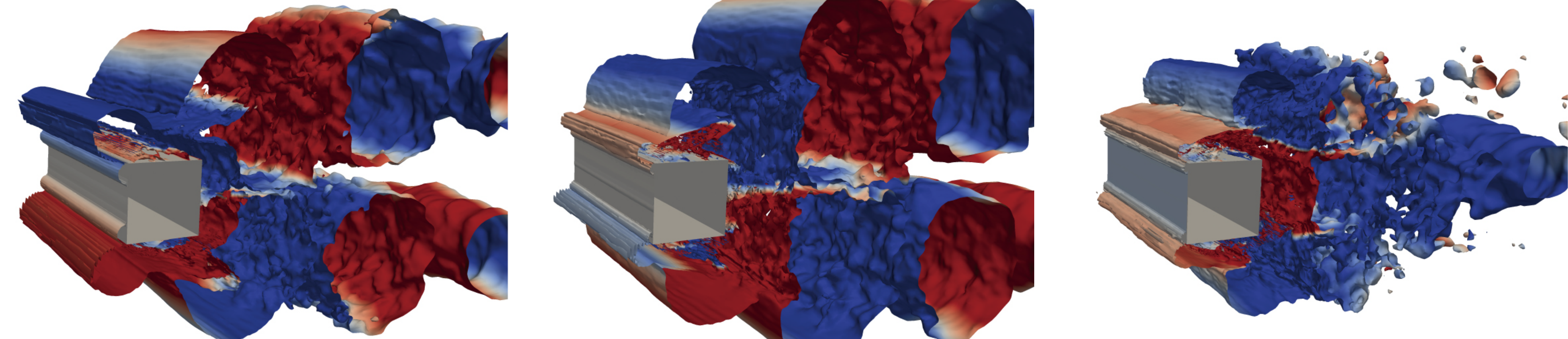}
        \put(0,20){\fig{a}}
    \end{overpic}
    \hfill
    \\ \vspace{1em}
    
    \hfill
    \begin{overpic}[width=1\textwidth]{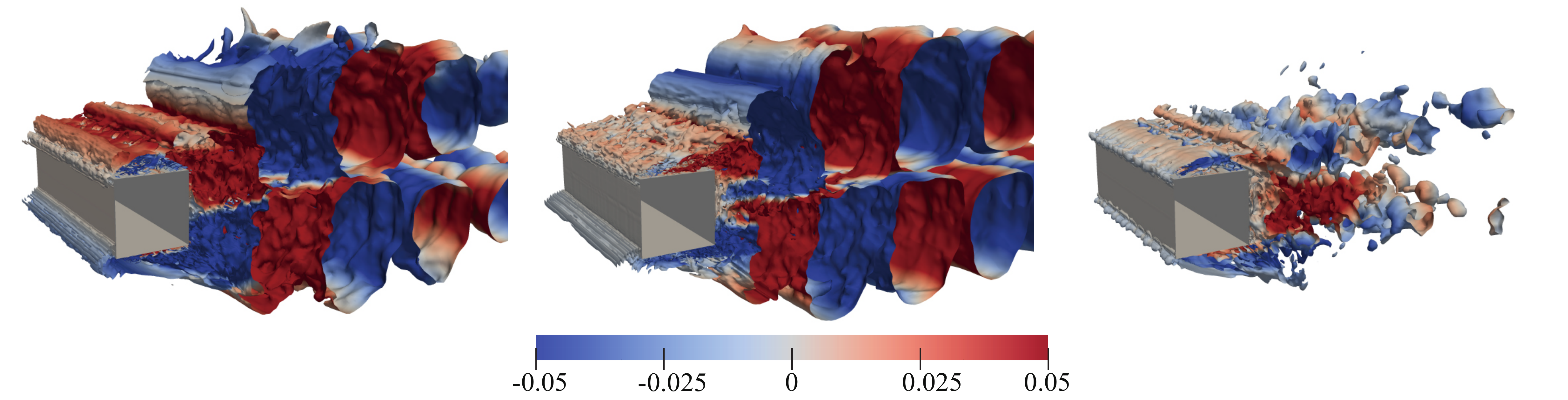}
        \put(0,23){\fig{b}}
    \end{overpic}
    \hfill
    % \\ \vspace{1em}
    
    % \hfill
    % \begin{overpic}[width=0.3\textwidth]{Figures/pod_bar.pdf}
    %     % \put(3,50){\fig{b}}
    % \end{overpic}
    % \hfill
\caption{Streamwise velocity components of the first three POD modes for: \fig{a} baseline ﬂow, \fig{b} ﬂow with the SAC control.} 
\label{fig:velocity_pod}
\end{figure}

% 为了理解吹吸气控制器的机制，我们基于波动动能对baseline和受控流进行了 POD 分析，以说明它们在非稳定流结构上的差异。Fig. \ref{fig:volicity_pod} 显示了baseline与DRL控制流在前三个 POD 模式的流向速度分量。在非强制流中，涡流在方柱前端开始发展，直到流向后缘形成大涡，且明显呈现对称分布。经过DRL控制后，可以观察到，涡流在方柱前缘被吹吸气吸入，对应着Fig. \ref{fig:actioncl}中$Q_1$ 和 $Q_2$ 的吸气动作。涡流在方柱表面不再卷曲成大涡，后缘吹吸气$Q_3$ 和 $Q_4$进行吹气将涡流推离，增大方柱尾流再循环区面积。因此在三阶模态下的对称涡流大小相比于非强制流都有所减少。总之，DRL射流吹吸气器延迟了主要旋涡的形成，从而进一步减少了这些旋涡对方柱造成的压力波动。
To comprehend the blow-suction controller mechanism, proper orthogonal decomposition (POD) analyses are conducted based on fluctuating kinetic energy for both baseline and controlled flows to delineate distinctions in their unsteady flow structures. Fig. \ref{fig:velocity_pod} illustrates the streamwise velocity components for the baseline and SAC controlled flows, focusing on the first three POD modes. In the unforced flow, vortices initiate development at the leading edge of the square cylinder, culminating in the formation of a large vortex at the trailing edge and exhibiting a symmetric distribution. After control, it becomes evident that the vortex undergoes suction at the leading edge of the square cylinder, corresponding to the actions of $Q_1$ and $Q_2$ depicted in Fig. \ref{fig:actioncl}. Consequently, the vortex no longer coils into a large structure on the surface of the square cylinder. Moreover, the blowing and suction actions of $Q_3$ and $Q_4$ at the trailing edge push the vortex away, expanding the recirculation zone in the wake region of the square cylinder. Consequently, the symmetric vortex size across the three modes is diminished compared to the unforced flow. In conclusion, the DRL jet blowing and suction actuator postpones the formation of major vortices, thereby further mitigating pressure fluctuations induced by these vortices in the square cylinder.

\subsection{Flow control results using the SDTL-SAC algorithm} \label{sec:Results2}

%FIG TL训练示意图 
\begin{figure} \centering
    \begin{overpic}[width=0.9\textwidth]{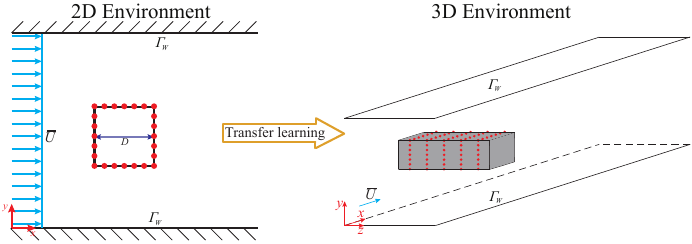}
    \end{overpic}
    \hfill
    \caption{Process of transfer learning under state dimension mismatch from 2D square cylinder flow field to 3D square cylinder environments.} 
    \label{fig:2dto3d}
\end{figure}
% 介绍2dto3d的过程
% As shown in Fig. \ref{fig:volicity_pod},为了达到2D流场向3D流场进行迁移学习的任务. 2D方柱流场在表面均匀布置了24个压力探针, 而3D流场则布置了五圈,每圈24个压力点,目的是充分的探知流场信息. 且两个方柱案例均在方柱四角设置吹吸气孔进行控制. SDTL-SAC的目标是在状态空间不匹配的情况下进行迁移学习的训练,以达到加速收敛的效果. 
% The distribution of 24 pressure probes on the surface is uniform for the 2D square cylinder flow field, while for the 3D flow field, the pressure probes are organized in five circles, ensuring a comprehensive detection of flow field information. 
As depicted in Fig. \ref{fig:2dto3d}, in order to facilitate transfer learning from a 2D flow field to a 3D flow field, 24 pressure probes are uniformly deployed on the surface of the 2D square cylinder flow field. Additionally, 24 pressure probes in each of the five circles are arranged to comprehensively explore the flow field information in the 3D flow field. In both square cylinder cases, flow control is achieved through four jets located at the corners of the square cylinder. The objective of SDTL-SAC is to expedite the convergence of transfer learning in the presence of state space mismatch problems.

%FIG Re=1000的情况  cd cl  流场等
\begin{figure} \centering
    \begin{overpic}[width=0.48\textwidth]{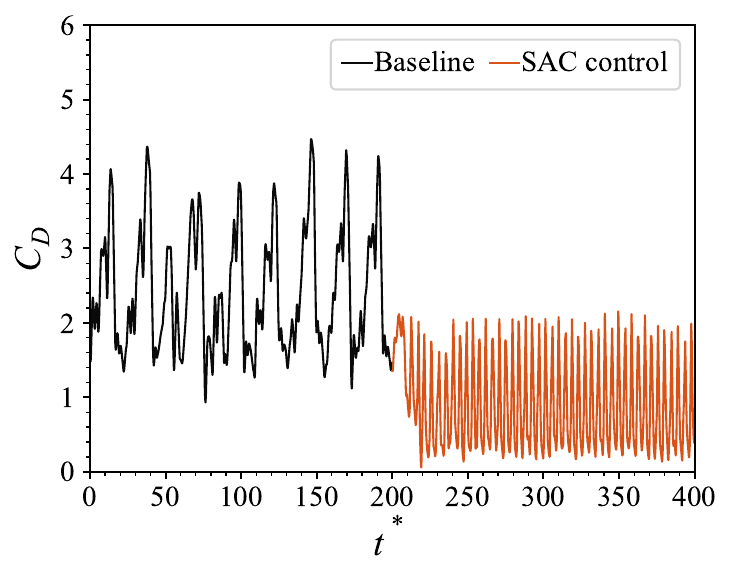}
        \put(0,75){\fig{a}}
    \end{overpic}
    \hfill
    \begin{overpic}[width=0.48\textwidth]{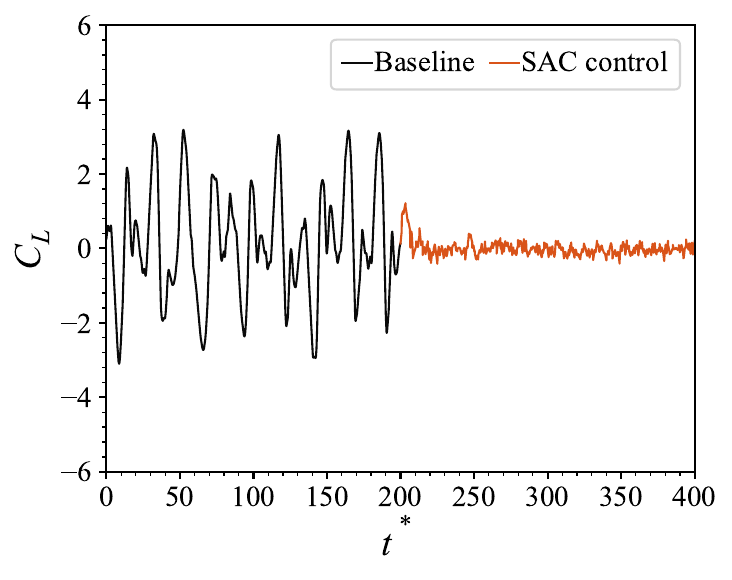}
        \put(0,75){\fig{b}}
    \end{overpic} 
\caption{\fig{a} Evolution of $C_D$ for the 2D square cylinder without (no jet) and with SAC control at $Re = 1000$; \fig{b} Temporal variations in smoothed $C_L$ for the square cylinder without (no jet) and with SAC control at $Re = 1000$.} \label{fig:1000cdcl}
\end{figure}
% 该吹吸气设置在1000的效果
% As shown in Fig. \ref{fig:1000cdcl}, 在经过大量的DRL训练后, 该吹吸气设置在Re=1000的情况下同样取得了较好的减阻降升力效果, 减阻比例为63.2\%,而阻力和升力波动的下降比例分别为30.3\%,  93.1\%. 这说明使用四角吹吸气的方式, 在2D方柱上有较好的效果.
Control results of this blowing and suction configuration in 2D square cylinder flow field as illustrated in Fig. \ref{fig:1000cdcl}, extensive DRL training of the blowing and suction configuration resulted in significant drag and lift reduction performance at the condition of 2D flow field and $Re=1000$. Specifically, the drag reduction ratio reached 63.2\%, accompanied by a reduction of 30.3\% in drag fluctuation and an impressive 93.1\% reduction in lift fluctuation. These outcomes underscore the efficacy of the four jets blowing and suction policy, particularly in the flow field of the 2D square cylinder. This control performance lays the foundation for transfer policy to a 3D flow field of the square cylinder.

%FIG 迁移到22000的cd cl reward对比
\begin{figure} \centering
    \begin{overpic}[width=0.48\textwidth]{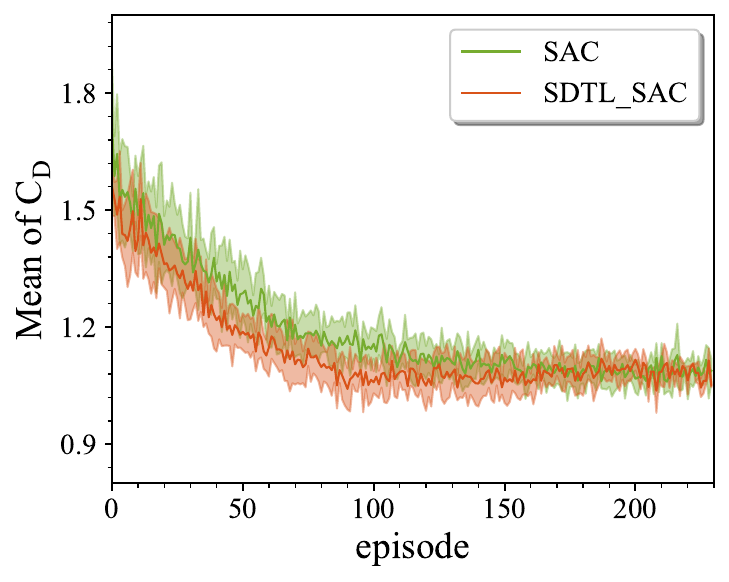}
        \put(0,75){\fig{a}}
    \end{overpic}
    \hfill
    \begin{overpic}[width=0.48\textwidth]{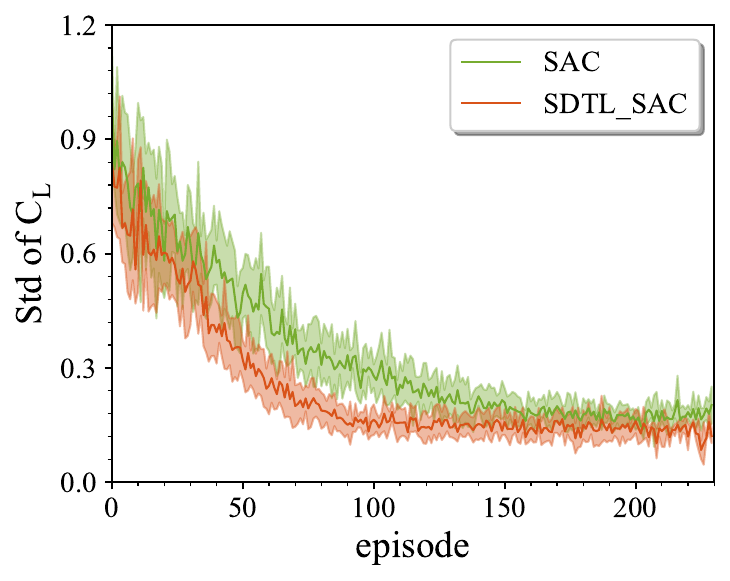}
        \put(0,75){\fig{b}}
    \end{overpic}
    \\
    \begin{overpic}[width=0.48\textwidth]{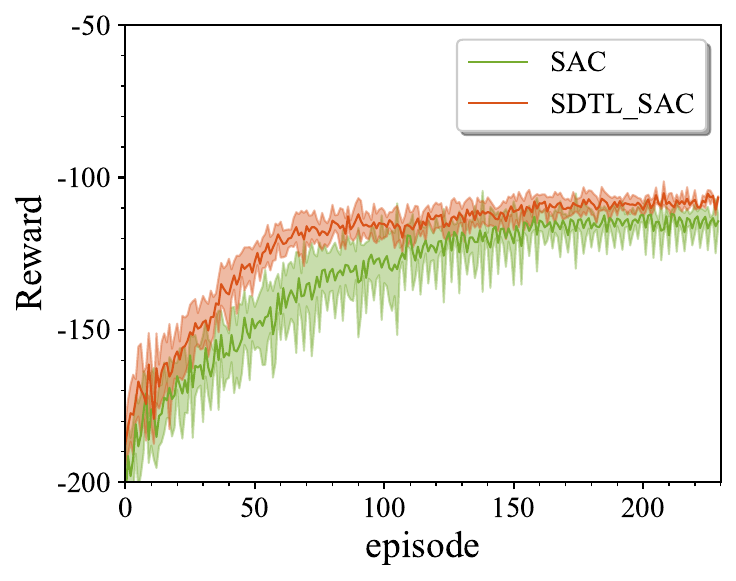}
        \put(0,75){\fig{c}}
    \end{overpic}
        \hfill
    \begin{overpic}[width=0.48\textwidth]{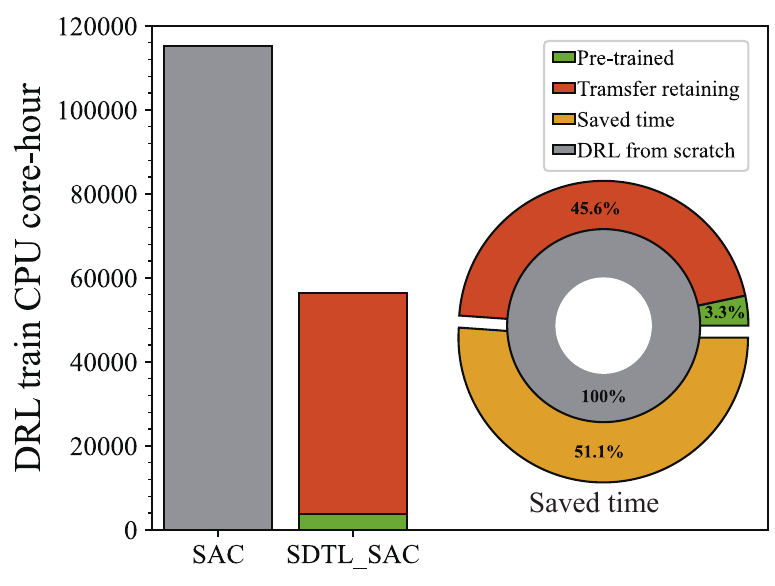}
        \put(0,75){\fig{d}}
    \end{overpic}
    \caption{Evolution of \fig{a} mean drag reduction, \fig{b} reward, and \fig{c} std of lift coefficient with basic SAC and SDTL-SAC method, respectively; \fig{d} shows the computational resources saved by SDTL-SAC method.
    } 
    \label{fig:reward}
\end{figure}
%reward比较   需要重写
% 图（ref{fig:reward}\fig{a}）显示了随着DRL训练集数量的增加，平均C_D$的变化情况，比较了基本SAC算法和SDTL-SAC算法的训练条件。两种算法都在十个环境中进行并行训练，与基线算法相比，阻力大幅降低。具体来说，与基线相比，两种算法 SAC 和 SDTL-SAC 的平均 $C_D$ 分别降低到 1.06 和 1.03。然而，从图中的奖励学习曲线可以看出明显的区别。SDTL-SAC 方法的奖励曲线在整个过程中呈现出较快的上升趋势，最终达到 -106 左右。相比之下，SAC 算法在第 50 次和第 180 次之间会遇到不稳定的情况，表现为显著的波动，并在训练完成后出现一个相对较低的值。造成奖励值增长率差异的另一个因素是 $C_L$ 的标准偏差，如图所示。SAC 算法将升力波动降低到了 0.20，而 SDTL-SAC 方法则迅速抑制了方形圆柱体的升力波动，在第 90 集时达到了 0.13。SDTL-SAC 方法表明，它能够在更短的时间内达到更高的控制水平。
% 我们注意到，我们的算法（SDTL）优于vanillaSAC算法，在训练的早期阶段都获得了更高的收益，最终性能也高得多。这证明：首先，这些任务确实具有结构上的共同性，因此，在一项任务中训练有素的教师策略可用于加速不同任务的学习；其次，SDTL 是实现这种知识转移的成功方法。即使教师和学生的 MDP 具有不同的状态和行动空间，SDTL 也能奏效，它是通过任务对齐的学习嵌入和互信息损失优化来实现的。
Fig. \ref{fig:reward}\fig{a} illustrates the variation in mean $C_D$ with an increasing number of DRL training sets, comparing the training conditions of basic SAC and SDTL-SAC algorithms. Both algorithms undergo parallel training in 10 environments \citep{wang2023deep}, demonstrating substantial drag reduction compared to the baseline. Specifically, compared with the baseline, the mean $C_D$ of the two algorithms SAC and SDTL-SAC is reduced to 1.06 and 1.03 respectively. Another contributing factor to the difference in reward value growth rate is the standard deviation of $C_L$, as depicted in Fig. \ref{fig:reward}\fig{b}. While the SAC algorithm reduces lift fluctuation to 0.20, the SDTL-SAC method promptly suppresses square cylinder lift fluctuation, reaching 0.13 by episode 90. The SDTL-SAC method exhibits the ability to attain a superior control level within a shorter timeframe when confronted with state dimension mismatch during transfer learning.

%FIG   提升多少速度对比
However, notable distinctions emerge from the reward learning curve presented in Fig. \ref{fig:reward}\fig{c}. The reward curve of the SDTL-SAC method shows a faster upward trend throughout the process, eventually reaching around -106. In contrast, the SAC algorithm encounters instability between episodes 50 and 180, marked by significant fluctuations followed by a relatively low value after training completion. Additionally, the reward curve can serve as a criterion for assessing the convergence of DRL training. It is evident that the SDTL-SAC method significantly accelerates DRL training progress. Specifically, basic SAC training from scratch requires 160 episodes to achieve convergence, while training a 2D square cylinder pre-trained policy with SDTL-SAC method only requires approximately 70 epochs to do so. 

Fig. \ref{fig:reward}\fig{d} illustrates the computational resources efficiency of the SDTL-SAC method. The basic SAC training demands substantial computational time and resources. In the case of computational resource consumption of the SDTL-SAC method, the training of the 2D control policy accounts for only 3.3\% of the total computational resources. The transfer retaining phase, meaning transfer learning in the 3D flow field with the pre-trained model, occupies 45.6\% of the total computational resources. In addition, the SDTL-SAC method achieves a noteworthy reduction in computational time, reducing 51.1\% of the original training time. This highlights the significant time savings afforded by the SDTL-SAC method in DRL training, even under state dimensions mismatch.

%FIG 展示actor_current_ps critic1等参数的变化图
\begin{figure} \centering
    \begin{overpic}[width=0.48\textwidth]{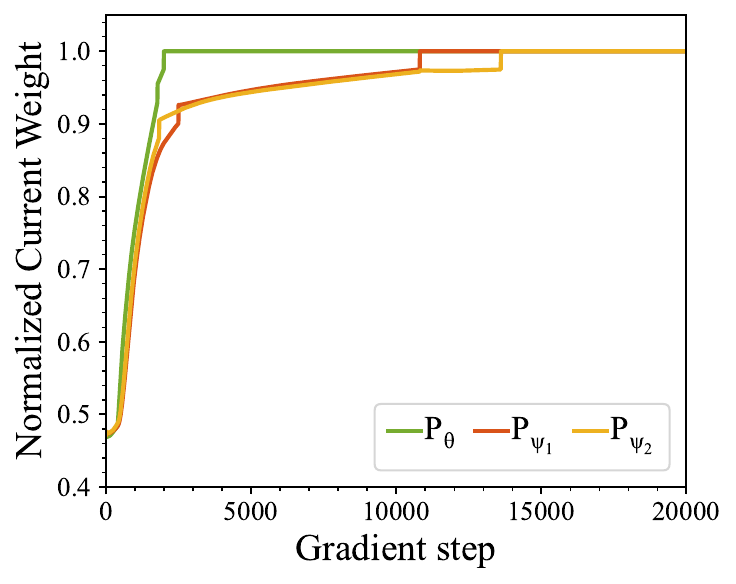}
        % \put(0,75){\fig{a}}
    \end{overpic}
    \hfill
    \caption{The value of the weighting assigned to the current policy representations, denoted as $p_\theta, p_{\psi_1}, p_{\psi_2}$, within the context of the weighted linear combination is explored during the transfer process of a 2D square cylinder control policy to 3D square cylinder environments.} 
    \label{fig:chaocanshu}
\end{figure}
%超参数变化
%  As shown in Fig. \ref{fig:chaocanshu}，我们绘制了与教师进行加权线性组合时，当前表示的权重值。我们观察到，在训练的初期，训练的策略学会信任自己学到的表征,而不是吸收来自不同老师的知识. 
% 另外,经过\citep{guastoni2023deep}验证,这种跨域迁移学习的关键在于当前策略与源策略之间的任务相似性. 使用不同类型(非方柱)的源策略进行迁移学习时,可能达不到本文迁移学习的效果. 
As illustrated in Fig. \ref{fig:chaocanshu}, the learnable mixing weights $p_\theta, p_{\psi_1}, p_{\psi_2}$ undergo evolution over time. As the training process advances, these weights progressively converge toward 1. In the initial stages of training, the current policy gradually assimilates valuable representations from the source policy until achieving complete independence from the source policy. In addition, the success of cross-domain transfer learning hinges on the task similarity between the current and source policy. 

\section{Conclusions and outlook} \label{sec:Conclusions}
%第一段
% 本研究介绍了利用位于方柱体上的表面压力传感器作为 DRL 代理的唯一输入，在实际主动流控制方面取得的重大进展。并应用state dimensional dismatch(SDTL-SAC)方法进行2D到3D流场的迁移学习训练。得到了明显的加速效果。这种方法有可能推动深度强化学习在现实世界中的应用，如减少高层建筑的阻力和升力。本研究的主要成果总结如下。
This study represents a notable advancement in practical AFC by utilizing a surface pressure sensor situated on a square cylinder as the sole input to the DRL agent. The study focuses on transfer learning from 2D to 3D square cylinder flow fields in the state dimensional mismatch by employing the proposed SDTL-SAC method, resulting in significant acceleration outcomes. This method holds promising implications for enhancing DRL applications in real-world scenarios, particularly in the reduction of drag and lift forces for structures like high-rise buildings. The primary findings of this study are succinctly outlined as follows.

%第二段
% 首先，我们在3D高雷诺数（$Re$ =22000）方柱主动控制的数值模拟中实施了 DRL。使用DRL算法与多吹吸气控制后，阻力系数的降低非常显著，阻力系数降低高达52.3\%。且$C_D$与$C_L$ 的波动几乎被完全抑制. 在控制过程的最后阶段，代理保持相当大的动作幅度，以有效扰乱方柱两侧的涡流。因此，附着在表面的漩涡会延迟脱落，从而减少对方柱产生的升力冲击。这种精确的自适应控制策略使得多吹吸气射流具有更优越的流量控制性能。此外，从时间平均涡度等值线观察到的效果表明，多吹吸气射流具有有效管理方形圆柱体流场涡流脱落模式的潜力。突出表明了这种方法在改善空气动力性能方面的功效。
First, the basic SAC algorithm is arranged to numerically simulate active control of a 3D square cylinder at $Re = 22000$. The utilization of the DRL algorithm specifically with multiple jets control, led to a substantial reduction in $C_D$ reaching an impressive 52.3\%. Concurrently, fluctuations in both $C_D$ and $C_L$ are suppressed by 74\% and 91.7\%, respectively. During the last phase of the control process, the agent maintained a significant action amplitude, effectively disrupting the vortices on both sides of the square cylinder. Consequently, the vortices adhered to the square cylinder surface and experienced delayed shedding, contributing to a reduction in lift impact generated by the square cylinder. This precise adaptive control strategy demonstrated superior flow control performance for the jet configuration. Additionally, the mean velocity contours illustrated that the blowing and suction multiple jets have the potential to effectively manipulate vortex shedding patterns in the flow field of square cylinders. The effectiveness of this approach in enhancing aerodynamic performance is underscored.

%第三段
% 在保持性能的同时，还对DRL训练速度进行了优化。我们提出一种DRL中的迁移学习算法SDTL-SAC，其中current和source代理可以有任意不同的状态空间。我们通过学习一个编码器来产生嵌入，从source网络中提取有用的表示来实现这一点。并在2D到3D方柱的迁移任务上进行的实验表明. SDTL-SAC算法可以加快高雷诺数方柱的训练过程.此外，研究还表明，TL可以带来更稳定的决策，这对流量控制具有潜在的益处。
Furthermore, the training speed of DRL is enhanced while preserving performance by introducing the State dimensional transfer Learning with the soft actor-critic algorithm (SDTL-SAC). This transfer learning algorithm accommodates situations where the current and source agents possess arbitrarily different state dimenshion. The key innovation involves training an encoder to generate embeddings and extracting valuable representations from the source network. Experiments on a transfer task from 2D to 3D square cylinder flow field, demonstrate that the SDTL-SAC algorithm accelerates the training process for high $Re$ square cylinders which can lead to up to a 51.1\% reduction in training time. Additionally, the results indicate that transfer learning contributes to more stable decision-making, potentially enhancing its utility in flow control applications.

%第四段
% 迁移学习是一个非常热门的研究领域，针对不同的流体动力系统选择合适的迁移学习方法，可以利用更少的时间达到最佳的控制水平\citep{Brunton2016pnas,Loiseau2018jfm}。处理这些特征并将其用作 DRL 状态是一种很有前景的方法。进一步扩展该方法的应用为在不同状态与动作空间内做迁移学习。尽管实现的任务更加复杂，但也是一种更现实的设置。 SDTL-SAC方法是在不同流场间建立迁移学习的有效办法，同时在高雷诺数下实现了最优的$C_D$和$C_L$降低性能，为驯服复杂的流体动力学系统提供了一条前景广阔的途径。% 通过将已有的二维和三维流动控制研究与迁移学习相结合，我们期望提高对真实流场行为的理解，并为更有效地实现流场控制提供新的途径。这项研究的意义在于推动流动控制领域的发展，为工程应用和科学研究提供更为可靠和高效的解决方案。
Transfer learning is a widely explored research domain, and the judicious selection of transfer learning methods tailored to specific fluid dynamic systems can significantly enhance control efficiency within a reduced timeframe. Leveraging these features and incorporating them as states in DRL holds considerable promise. A natural progression of this methodology involves extending its application to accommodate transfer learning across distinct state and action spaces, introducing a more realistic yet inherently complex implementation challenge. The SDTL-SAC method emerges as a robust approach for transfer learning across diverse flow fields, yielding optimal reductions in $C_D$ and $C_L$ at elevated Reynolds numbers. This method holds the potential for effectively managing intricate fluid dynamic systems. By integrating existing research on 2D and 3D flow control with transfer learning, we aim to enhance the understanding of the behavior of real flow fields and pave the way for more effective implementation of flow field control. The significance of this study lies in its contribution to advancing the field of flow control, offering more reliable and efficient solutions for engineering applications and scientific research.

\section*{Acknowledgement}
This study is supported by National Natural Science Foundation of China (52278493, 52108451), Shenzhen Science and Technology Program (KQTD20210811090112003). This work is also supported by National Science Foundation of China (NSFC) through grants 12172109 and 12172111.

\appendix
%附录A 放出code
%附录C
%表格展示使用参数
\section{Hyperparameters} \label{sec:hyperparameters}
Table \ref{tab:Hyperparameters_flow} presents the main numerical parameters of both the simulated case and the learning algorithm.

\begin{table}
    \centering
    \caption{Configurations of flow simulation} \label{tab:Hyperparameters_flow}
    \renewcommand\arraystretch{1}
        \begin{tabularx}{\textwidth}{lc}
        \toprule
        \textbf{Parameter} & \textbf{Value} \\
        \midrule
        \multicolumn{2}{c}{\textbf{Flow simulation set-up}} \\ [0.1cm]
        Numerical time step (non-dimensional $dt$)   &  $5\times10^{-5}$  \\
        Maximum action amplitude (non-dimensional)   & 3    \\
        Control step length   & 100 $dt$   \\
        \multicolumn{2}{c}{\textbf{SAC hyperparameters}} \\ [0.1cm]
        Actor architecture   &  $512\times 256$ (two fully connected layers) \\
        Critic architecture   & $512\times 256$ (two fully connected layers) \\
        Actor leaning rate  &  $ 3\times10^{-4}$ \\
        Critic leaning rate  &  $ 2\times10^{-4}$ \\
        Discount factor  & 0.97 \\
        Alpha  & 0.5 \\
        Optimiser &  Adam\citep{kingma20153rd} \\
        \multicolumn{2}{c}{\textbf{SDTL-SAC hyperparameters}} \\ [0.1cm]
        Parallelized environments  &  10 \\
        Number of action per episode & 80 \citep{wang2022drlinfluids} \\
        CPU time per episode   & 2.5h  \\
        % Total CPU time of training stage   &  $\approx$ 6250 CPUh \\  % 72h一个算例  69个算例算例
        \bottomrule
        \end{tabularx}
\end{table}

%附录B SAC+SDTL
%伪代码示意表格
\section{State dimension mismatch transfer learning.} \label{sec:SDTL-SAC}
% 算法流程
Algorithm \ref{alg:SDTL-SAC} describes the steps of SDTL-SAC.

\SetKwComment{Comment}{/* }{ */}
\SetKwInOut{init}{Initialization}
\begin{algorithm}[htb]
\caption{SDTL algorithm with Soft Actor-Critic.}\label{alg:SDTL-SAC}
% \KwIn{fixed source policy and value networks ${\theta'},{\psi'_i}$, current policy and value networks $\theta,\psi_i $, policy and value encoder parameters $\phi_\pi,\phi_i$, policy and value variational distribution parameters $\omega_\pi,\omega_i$, set of coupling parameters for policy and value networks $\left\{p_\pi\right\},\left\{p_i\right\}$
\KwIn{${\theta'},{\psi'_i}$: fixed source policy and value networks ${\theta'},{\psi'_i}$\\
$\theta,\psi_{i} $: current policy and value networks\\ 
$\phi_\pi,\phi_{i}$: policy and value state encoder parameters\\
$\omega_\pi,\omega_{i}$: policy and value variational distribution parameters\\
$\left\{p_\pi\right\},\left\{p_i\right\}$: set of coupling parameters for policy and value networks 
}
\init{target network weights $\overline{\theta_i} \gets \theta_i$, replay buffer $\mathcal{D} \gets \emptyset$}
\For{each iteration}{
    \For{each environment step}{
        $a_t \sim \pi_\phi\left(a_t \mid s_t\right)$\;
        $s_{t+1} \sim p\left(s_{t+1} \mid s_t, a_t\right)$\;
        $\boldsymbol{S}_t \gets \boldsymbol{g}(\{s_1,s_2,\dots,s_{t+1}\} \cup \{a_1,a_2,\dots,a_{t}\})$\;
        $\mathcal{D} \leftarrow \mathcal{D} \cup \left\{(\boldsymbol{S}_t, a_t, r(s_t, a_t))\right\}$\;
    }

    \For{each gradient step}{
        Update $\psi_i$ with $\lambda_{Q} \hat{\nabla}_{\psi_i}L_{Q}(\psi_i,\phi_i,{\psi'_i}) \mathrm{for~}i \in {1,2}$\;
        Update $\phi_i$ with $\hat{\nabla}_{\phi_i} \big[L_{Q}(\psi_i,\phi_i,{\psi'_i})+L^{Q}_{\mathrm{MI}}(\phi_i,\omega_i) \big] \mathrm{for~}i \in {1,2}$\;
        Update $\omega_i$ with $\hat{\nabla}_{\omega_i} L^{Q}_{\mathrm{MI}}(\phi_i,\omega_i) \mathrm{for~}i \in {1,2}$\;
        Update $\left\{p_\psi\right\}$ using $\left[L^{Q}_{\mathrm{coupling}}+L_{Q}\right] $
                
        Update $\theta$ with $\lambda_\pi \hat{\nabla}_\theta L_\pi(\theta,\phi_\pi,\theta')$\;
        Update $\phi_\pi$ with $\hat{\nabla}_{\phi_\pi} \big[L_\pi(\theta,\phi_\pi,\theta')+L^\pi_{\mathrm{MI}}(\phi_\pi,\omega_\pi) \big]$\;
        Update $\omega_\pi$ with $\hat{\nabla}_{\omega_\pi} L^\pi_{\mathrm{MI}}(\phi_\pi,\omega_\pi)$\;
        Update $\left\{p_\pi\right\}$ using $\left[L^\pi_{\mathrm{coupling}}+L_\pi\right]$
        
        Update $\overline{\theta_i}$ and $\alpha$ according to \citep{haarnoja2018soft}\;

    }
}
% \KwOut{actuator control vector $a \in\mathbb{R}^{1\times j}$}
\end{algorithm}

\bibliographystyle{jfm}
\bibliography{jfm}

\end{document}